\UseRawInputEncoding
\documentclass[
article,
aps,
preprint,    
showpacs,
preprintnumbers, 
 amsmath,amssymb,
]{revtex4-2}
\usepackage{amsfonts}
\usepackage{booktabs}
\usepackage{siunitx}
\usepackage{graphicx}
\usepackage{dcolumn}
\usepackage{bm}
\usepackage{lineno}
\usepackage[version=3]{mhchem}
\usepackage{color}
\usepackage{rotating}

\begin{document}
\title{Band lineup at hexagonal Si$_x$Ge$_{1-x}$/Si$_y$Ge$_{1-y}$ alloy interfaces}%
\author{  Abderrezak Belabbes$^{1}$}
\email {abderrezak.belabbes@uni-jena.de}
\author{ Silvana Botti$^2$}
\author{ Friedhelm Bechstedt$^2$}
\affiliation{%
$^1$ Department of Physics, Sultan Qaboos University, P.O. Box 36, PC 123, Muscat, Oman \\
$^2$  Institut f\"ur Festk\"orpertheorie und -optik, Friedrich-Schiller-Universit\"at Jena, Max-Wien-Platz 1, 07743 Jena, Germany
}%
\begin{abstract}
The natural and true band profiles at heterojunctions formed by hexagonal Si$_x$Ge$_{1-x}$ alloys are investigated by a
variety of methods: density functional theory for atomic geometries, approximate quasiparticle treatments for electronic
structures, different band edge alignment procedures, and construction of various hexagonal unit cells to model alloys and heterojunctions.
We demonstrate that the natural band offsets are rather unaffected by the choice to align the vacuum level or the branch point energy, as well as by the use of a hybrid or the Tran-Blaha functional. At interfaces between Ge-rich alloys we observe a type-I heterocharacter with direct band gaps, while Si-rich junctions are type-I but with an indirect band gap. The true band lineups at pseudomorphically grown heterostructures are
strongly influenced by the generated biaxial strain of opposite sign in the two adjacent alloys. Our calculations show that the type-I character of the interface is reduced by
strain. To prepare alloy heterojunctions suitable for active optoelectronic applications, we discuss how to decrease the
compressive biaxial strain at Ge-rich alloys.
\end{abstract}
\maketitle

\section{\label{sec1}Introduction}

Silicon (Si) grown in the diamond structure is the key material of modern micro- and nano-electronic devices and integrated
circuits. However, it is an indirect semiconductor, like other group-IV elements as germanium (Ge). Active optoelectronic
devices, such as light-emitting diodes and lasers, cannot be produced using Si and Ge \cite{Soref:1993:InProc,Zimmermann:2010:Book}.
Therefore, optical intrachip communication cannot be realized within the Si-based CMOS technology, that avoids the
combination with III-V compounds, because of possible unintentional doping.

The semiconductor Ge possesses a direct gap only slightly above the indirect gap energy \cite{NSMArchive:html}. Various strategies to make Ge a direct-gap semiconductor, therefore  suitable for Ge-on-Si optoelectronics, have been explored \cite{Liu.ea:2012:TSF}. One favored manipulation of Ge is the
application of uniaxial or biaxial strains
\cite{Zhang.Crespi.ea:2009:PRL,Saladukha.Clavel.ea:2018:PRB,Jung.ea:OE:2021,Mellaerts.Afanasaev.ea:2021:AAMI}. Another
approach is the use of hexagonal polytypes $n$H ($n=2,4,6$) \cite{Raffy.Furthmueller.ea:2002:PRB}, in particular the 2H crystal structure, also called lonsdaleite, of Si \cite{Roedl.Sander.ea:2015:PRB} and Ge \cite{Roedl.Furthmueller.ea:2019:PRM}. Indeed, these latter crystals can be grown in form of nanowires \cite{Hauge.Verheijen.ea:2015:NL,Dixit.Shukla:2018:JoAP,Fadaly:2020:N}. Crystalline 2H-Ge has a direct fundamental gap, but is a pseudodirect semiconductor, since its lowest-energy optical transitions only show a negligible
dipole strength \cite{Roedl.Furthmueller.ea:2019:PRM}. Tensile uniaxial strain parallel to the $c$-axis of the lonsdaleite geometry
is predicted to lead to a conduction band inversion and, therefore, to strong optical transitions \cite{Suckert.Roedl.ea:2021:PRM}. In general,
structural and chemical perturbations to the ideal bulk crystal tend to make 2H-Ge suitable for optoelectronic applications \cite{Belabbes.Bechstedt.ea:2021:PSSRRL}.
Indeed, alloying hexagonal Ge with Si to 2H-Si$_x$Ge$_{1-x}$ gives rise to strong photoluminescence, whose photon energy shifts
toward higher energies  with increasing Si composition \cite{Fadaly:2020:N}. Both effects have been also predicted theoretically
\cite{Cartoixa.Palummo.ea:2017:NL,Borlido.Suckert.ea:2021:PRM}.

The energy-light conversion in light-emitting diodes (LEDs) and lasers can be improved by quantum confinement effects, e.g. in
quantum-well LEDs and lasers \cite{Weisbuch.Nagle:2007:PS,Chuang:1995:Book}. In addition, quantum confinement leads
to a decrease of the emission wavelength. Confinement and its effect depend on the band alignment at interfaces, that can produce discontinuities
between well and barrier materials. Moreover, during pseudomorphic growth, due to the different lattice constants,
biaxial strain is induced in the quantum well system, which also modifies the quantum confinement and the light emission.
The combination of these effects may also help to increase the light emission efficiency of Si$_x$Ge$_{1-x}$ alloy systems. However, its
understanding and control require the detailed knowledge of band discontinuities $\Delta E_c$ and $\Delta E_v$ of the conduction
band minimum (CBM) and valence band maximum (VBM), respectively, at the interface of two alloys forming the heterostructure, namely 2H-Si$_x$Ge$_{1-x}$
and 2H-Si$_y$Ge$_{1-y}$ ($x<y$), resulting in so-called band lineups. To this end, the chemical effects due to the different Si contents $x$ and $y$ and the strain effects due to biaxial growth have to be studied. For the corresponding diamond Si and Ge
alloys, theoretical studies of band discontinuities have been carried on several decades ago \cite{VandeWalle.Martin:1986:PRB}
and have been refined over the years \cite{VandeWalle.Neugebauer:2003:N,ElKurdi.Sauvage.ea:2006:PRB}. In the hexagonal
case, only in a very recent paper the first investigation for alloys has been proposed \cite{Wang.Zhang.ea:2021:APL}.

In this paper, we perform accurate \textit{ab initio} calculations to investigate hexagonal Si$_x$Ge$_{1-x}$/Si$_y$Ge$_{1-y}$ heterostructures. We combine density
function theory (DFT), that we use for the optimization of the atomic geometry, with approximate quasiparticle (QP) approaches based on DFT for the determination of the electronic structure. More specifically, we compare two state-of-the-art approaches for the calculations of band structures of solids. Moreover, we consider different band alignment methods and evaluate their performance. The resulting natural band discontinuities are analyzed and compared with values for the cubic case. The
occurrence of realistic pseudomorphic interfaces is simulated using short-range superlattices composed by two different alloys. The accompanying
biaxial strain is modeled by means of the calculated elastic constants and atomic distances in the constrained materials. At last, we present and compare two methods to compute band offsets based on the band deformation potentials.

\section{\label{sec2}Methods}
\subsection{\label{sec2a}Structural, elastic and electronic properties}

The atomic configurations, lattice constants and elastic coefficients are optimized with DFT as implemented in the Vienna
Ab-initio Simulation Package (VASP) \cite{Kresse.Furthmueller:1996:PRB,Kresse.Furthmuller:1996:CMS}, using the projector-augmented wave method \cite{Kresse.Joubert:1999:PRB} and a plane-wave cutoff of 500~eV. The Ge3$d$ electrons are treated as valence
electrons. The Perdew-Becke-Ernzerhof XC functional PBEsol revised for solids \cite{Perdew.Ruzsinszky.ea:2008:PRL} is applied. For hexagonal four-atom unit cells the Brillouin zone (BZ) integration is performed using  a
$\Gamma$-centered 12$\times$12$\times$6 ${\bf k}$-point grid. Atomic geometries are relaxed until the Hellmann-Feynman forces
are below 1~meV/{\AA}.

All electronic structure calculations include the spin-orbit interaction. The Kohn-Sham band structures \cite{Kohn.Sham:1965}
obtained using the DFT-PBEsol functional are known to significantly underestimate the band gap and  interband-transition energies, inducing consequent uncontrollable errors in the band offsets $\Delta E_c$ and $\Delta E_v$ \cite{Bechstedt:2015:Book}.
Accurate quasiparticle bandstructures are therefore required. To avoid the numerical effort associated to QP calculations within the GW approximation
\cite{Bechstedt:2015:Book}, we apply two approximate DFT approaches for quasiparticle band structures that are known to yield high-quality results~\cite{Borlido-Aull:JCTC-2019}. The first approach relies on the XC potential
of Tran and Blaha \cite{Tran.Blaha.ea:2007:JoPCM,Tran.Blaha:2009:PRL} that combines a modified Becke-Johnson (MBJ)
exchange potential \cite{Becke.Johnson:2006:TJoCP} with the local density approximation (LDA) correlation \cite{Kohn.Sham:1965} to give the MBJLDA functional.
The MBJLDA functional is known to yield excellent QP band energies around the fundamental gap for cubic and hexagonal Si and Ge crystals
\cite{Roedl.Furthmueller.ea:2019:PRM,Suckert.Roedl.ea:2021:PRM,Belabbes.Bechstedt.ea:2021:PSSRRL,Borlido.Suckert.ea:2021:PRM,Laubscher.Kuefner.ea:2015:JoPCM}.
The second approach relies on the hybrid XC functional of Heyd,
Scuseria, and Ernzerhof (HSE06) \cite{Heyd.Scuseria.ea:2003,Heyd.Scuseria.ea:2006:JoCP}. The band energies resulting
for 2H-Ge agree very well with the MBJLDA results
\cite{Roedl.Furthmueller.ea:2019:PRM,Suckert.Roedl.ea:2021:PRM,Belabbes.Bechstedt.ea:2021:PSSRRL}.

\subsection{\label{sec2b}Alloys and superlattices}

Calculating band alignments for disordered-alloy supercells would imply to perform a large number of expensive total energy and electronic-structure calculations. To avoid this bottleneck, we simulate hexagonal
Si$_x$Ge$_{1-x}$ alloys, in first approximation, using ordered four-atom hexagonal unit cells Si$_n$Ge$_{4-n}$ ($n=0,1,...,4$).
Although this choice makes possible to directly study only three intermediate compositions $x=n/4$ ($n=1,2,3$), beside pure Si and pure Ge, it is a good compromise to speed up calculations and allow for the comparison of different methods. The atomic arrangements of the 4-atom cells with $n$ Si and $(4-n)$
Ge atoms ($n=0,...,4$) are equivalent by symmetry, apart from the case of Si$_2$Ge$_2$, for which three different bonding configurations, indicated with 
C1, C2 and C3 in Fig.~\ref{fig1}(a), have to be investigated. In C1 and C2 cells, SiGe bonds are stacked along the $[0001]$ direction so that Si and Ge atoms are aligned (C1), or alternate (C2). In the C3 cell the stacked bonds  consist of alternating Si-Si and Ge-Ge pairs. The structure C2 has the highest symmetry, because the vertical bonds in the $[0001]$ direction are always equal Si-Ge pairs, and is the most energetically favorable. While the lonsdaleite
systems Ge$_4$ and Si$_4$ possess the space group P6$_3$/mmc ($C^4_{6v}$), the symmetry is reduced to P3m1 ($C^1_{3v}$) for  all considered hexagonal Si$_x$Ge$_{1-x}$ ordered alloys, irrespective of the chosen inequivalent configuration of the Si$_2$Ge$_2$ alloy (see Fig.~\ref{fig1}(a)).
The structural, elastic, and energetic parameters needed to construct and discuss the electronic structures and the band
lineups between two SiGe alloys are listed in Table~\ref{tab1}. All the  values exhibit a monotonous variation in the composition range from pure
Ge to pure Si. Complete band structures of the Si$_n$Ge$_{4-n}$ cells obtained using the MBJLDA functional can
be found in Refs. \cite{Fadaly:2020:N,Belabbes.Bechstedt.ea:2021:PSSRRL,Weisbuch.Nagle:2007:PS}. For 2H-Ge and Si$_1$Ge$_3$
a comparison of HSE06 and MBJLDA band energies is presented in Refs. \cite{Roedl.Furthmueller.ea:2019:PRM,Belabbes.Bechstedt.ea:2021:PSSRRL}.

\begin{table}[h!]
\caption{Structural parameters (lattice parameters $a$, $c$, and $u$), elastic parameters (bulk modulus $B$, its pressure derivative $B'$, and biaxial strain modulus $Y$), and total energy $E_{\rm tot}$ for the seven chemically- or symmetry-inequivalent hexagonal
compounds Si$_n$Ge$_{4-n}$. The values of $Y$ are taken from Ref.~\cite{Borlido:unpubl}. In the case of Si$_2$Ge$_2$, values for the three atomic configurations in Fig.~\ref{fig1} are listed together with their arithmetic averages.
}
\centering
\footnotesize
\begin{ruledtabular}
\begin{tabular}{ccccccccl} 
configuration          & $a$ ({\AA}) & $c$ ({\AA}) & $u$ & $B$ (GPa) & $B'$ & $Y$ (GPa) & $E_{\rm tot}$ (eV/cell) \\ \hline
 \
Ge$_4$                     & 3.9960 & 6.5920 & 0.3743 & 68 & 4.76 & 171 &  0.000    \\
Si$_1$Ge$_3$         & 3.9464 & 6.5150 & 0.3742 & 74 & 4.69 & 189 &   \llap{$-$}0.760 \\ 
Si$_2$Ge$_2$ (C1) &  3.8944 &  6.4375            &   0.3740           & 81 & 4.51 & 203 & \llap{$-$}1.592 \\
Si$_2$Ge$_2$ (C2) & 3.9002             &  6.4376            & 0.3746 & 81 & 4.54 & 203 &\llap{$-$}1.606	\\
Si$_2$Ge$_2$ (C3) &   3.9002           &   6.4377           & 0.3737  & 81 & 4.52 & 201 & \llap{$-$}1.596\\
Si$_2$Ge$_2$ (average) & 3.8981 & 6.4376 & 0.3741& 81 & 4.52 & 202 & \llap{$-$}1.598   \\
Si$_3$Ge$_1$         & 3.8626 & 6.3828 & 0.3740 & 87 &4.38 & 216 &  \llap{$-$}2.398 \\
Si$_4$                       & 3.8264 & 6.3272 & 0.3739 & 94 & 4.24 & 225&\llap{$-$}3.249&  
\label{tab1}
\end{tabular}
\end{ruledtabular}
\end{table}

\section{\label{sec3}Natural band discontinuities}
\subsection{\label{sec3a}Vacuum level alignment}

Natural band discontinuities can be predicted from the bulk band structures of two semiconductors and/or insulators that constitute an heterointerface. Such a prediction, however, asks for a method to align the energy scales of the adjacent materials. The electron affinity rule \cite{Anderson:1962:SE,Bechstedt.Enderlein:1988:Book}
is based on the vacuum level alignment. The known electron affinities $A_1=-E^1_{c}$ and $A_2=-E^2_c$ of the two
nonmetals $j=1,2$ in contact, with $E^j_c$ as the lowest CBM measured with respect to the vacuum level, define
the conduction band discontinuity as
\begin{equation}\label{eq1}
\Delta E_c=A_1-A_2 \,.
\end{equation}
The QP gaps $E^j_g=I_j-A_j$ and the resulting ionization energies with respect to the vacuum level $I_j=-E^j_v$ yield the valence
band discontinuities as $\Delta E_v=I_2-I_1$.

The procedure to calculate band energies $E^j_c$ and $E^j_v$ with respect to the vacuum level \cite{Hoffling.Schleife.ea:2012:PRB}
is illustrated in Fig.~\ref{fig2}(a) for the examples of hexagonal polytypes Ge$_4$ and Si$_4$ but also Si$_n$Ge$_{4-n}$ alloys and the
cubic polytypes 3C-Ge and 3C-Si. First, the electrostatic potentials $V(z)$, averaged over the plane perpendicular to the 
$z$-axis, have to be extracted from the single-particle potential of the generalized Kohn-Sham equation. To define the vacuum level, the
tail of $V(z)$ at a surface is needed. We studied the $(0001)$ surfaces by investigating slabs of eight atoms, i.e., two hexagonal unit 
cells, in the $[0001]$ direction, separated by vacuum layers as thick as 12~{\AA}. The oscillations of $V(z)$ inside the material slabs 
are compared with the electrostatic potential of the bulk calculation. The comparison allows the determination of $E_c$, but
also $E^{ind}_c$, and $E_v$ with respect to the vacuum level (see Table~\ref{tab2}). The alignment gives rise to the band
lineups displayed in Fig.~\ref{fig2}(b).

\begin{table}[h!]
\caption{ Positions of band extrema,  $E_{v}$,  $ E_{c}$ and $ E_{c}^{\mathrm{ind}}$, as determined via the electron affinity rule 
for Si$_{x}$Ge$_{1-x}$(0001) surfaces. For comparison the values of 3C-Ge(111) and 3C-Si(111) are also listed. 
The natural band discontinuities $\Delta E_c=E^{\rm lower}_c-E^{\rm upper}_c$ \eqref{eq1} 
and $\Delta E_v=E^{\rm upper}_v-E^{\rm lower}_v$ as well as that for the indirect CBM $\Delta E^{\rm ind}_c$ are also given. Because of the
three configurations C1, C2, C3 and the average one, four offsets are displayed for either Si$_{0.25}$Ge$_{0.75}$/Si$_{0.5}$Ge$_{0.5}$ or 
Si$_{0.5}$Ge$_{0.5}$/Si$_{0.75}$Ge$_{0.25}$ heterostructures. All energies in eV.}
\footnotesize
\begin{ruledtabular}
\begin{tabular}{cccccccccc}
     Interface/alloy Si$_{x}$Ge$_{1-x}$ & Orientation& $E_{v}$ &  $E_{c}$   &    $E_{c}^{\mathrm{ind}}$  &  $\Delta E_{v}$   & $\Delta E_{c}$  & $\Delta E_{c}^{\mathrm{ind}}$   \\ 
 \hline

 \
3C-Ge                                                            &   (111)                & \llap{$-$}4.535	& \llap{$-$}3.879& \llap{$-$}3.910     &\llap{$-$}0.060 & \llap{$-$}0.283  &0.070  \\ 
2H-Ge                                                             & (0001)               &\llap{$-$}4.475         &\llap{$-$}4.162 &\llap{$-$}3.840       &0.056 & 0.198  & 0.043   \\
Si$_{0.25}$Ge$_{0.75}$                               &   (0001)                   &\llap{$-$}4.531      &\llap{$-$}3.964 &\llap{$-$}3.797      &0.152 & 0.317  &	0.029 \\
                                                                                                                                                                                                   &&&&&0.170 & 0.324 &\llap{$-$}0.023\\
Si$_{0.5}$Ge$_{0.5}(\mathrm{C}1)$  &       (0001)                  &\llap{$-$}4.683	           &\llap{$-$}3.647&\llap{$-$}3.768      &\llap{$-$}0.026 & 0.101 &\llap{$-$}0.053\\
Si$_{0.5}$Ge$_{0.5}(\mathrm{C}2)$  &      (0001)                   &\llap{$-$}4.701	           &\llap{$-$}3.640&\llap{$-$}3.820       &0.099 &	0.248  & \llap{$-$}0.015\\
Si$_{0.5}$Ge$_{0.5}(\mathrm{C}3)$  &      (0001)                   &\llap{$-$}4.505	           &\llap{$-$}3.863&\llap{$-$}3.850     &0.052&	0.033	  &  \llap{$-$}0.059 \\
Si$_{0.5}$Ge$_{0.5}(\mathrm{average})$  &      (0001)                 &\llap{$-$}4.630	   &\llap{$-$}3.716 &\llap{$-$}3.812          &0.034 &	0.026	  &  \llap{$-$}0.007 \\       
                                                                                                                                                                                                        &&&&&0.230 &	0.249 & 0.023 \\
                                                                                                                                                                                                          &&&&&0.105 &	0.102 & \llap{$-$}0.015	 \\ 
                                                                                                                                                                                                                                                                                                       
Si$_{0.75}$Ge$_{0.2	5}$  &   (0001)                                             &\llap{$-$}4.735  &\llap{$-$}3.614	&\llap{$-$}3.827      &0.142 &	0.449 & \llap{$-$}0.021\\
2H-Si                &      (0001)                                                                 &\llap{$-$}4.877	&\llap{$-$}3.165	&\llap{$-$}3.848     &0.343 & 0.195 & \llap{$-$}0.121    \\                  
3C-Si     &        (111)                                                       &\llap{$-$}5.220	&\llap{$-$}2.970&\llap{$-$}3.969  &&&& 
\label{tab2}
\end{tabular}
\end{ruledtabular}
\end{table}

For the alloy, the potential $V(z)$ at the surface depends on the surface termination, i.e., the distribution of the Si and Ge atoms in the four-atom cells and hence in the surface atomic layer. 
In comparison to the values in Table~\ref{tab2} and Fig.~\ref{fig2}(b), obtained for Si$_{1}$Ge$_{3}$ and a termination with the Si layer as the second one beneath the surface, other terminations induce small variations of 58 to 61 meV for $E_v$, $-$28 to 90 meV for $E_c$,
and $-$26 to 84 meV for $E_{c}^{\mathrm{ind}}$. These numbers prove for Si$_{0.25}$Ge$_{0.75}$ that, fortunately, the termination has a negligibly small influence
on the vacuum level position. Table~\ref{tab2} gathers several other interesting results. The ionization
energy $I=-E_v$ is rather constant for Ge-rich alloys. Its increase to the larger value of 2H-Si, and finally to 3C-Si, Table~\ref{tab2}, begins for alloys with at least a 50\% Si content. 

The absolute values of the ionization potential for the hexagonal allotropes, $I=4.48$~eV (2H-Ge) and $I=4.88$~eV (2H-Si), are however significantly smaller than the experimental values for the cubic
 phases $4.5-4.8$~eV (3C-Ge) and $5.10-5.33$~eV (3C-Si), respectively \cite{Bechstedt.Enderlein:1988:Book,Guo.Li.ea:2019:JPCC}. An increase
of the theoretical values to $I=4.54$~eV and $I=5.22$~eV occurs for 3C-Ge and 3C-Si, respectively, indicating not only the agreement
with measured ionization energies but also the predictive power of the MBJLDA functional. Other calculations using the HSE or other screened exchange XC functionals
for 3C-Si give rise to even larger values \cite{Hoffling.Schleife.ea:2012:PRB,Guo.Li.ea:2019:JPCC}. Full GW calculations deliver $I=5.46$~eV (3C-Si)
and $I=5.05$~eV (3C-Ge) when (111)2$\times$1 surfaces are investigated \cite{Hinuma.Grueneis.ea:2014:PRB}. The good agreement of measured and MBJLDA values for $I$ and $A$ confirm the validity of the vacuum level alignment for the determination of natural band offsets. The latter values 
$\Delta E_{v}$, $ \Delta E_{c}$ and $ \Delta E_{c}^{\mathrm{ind}}$ are also listed in Table~\ref{tab2}.

\subsection{\label{sec3b} Branch point alignment}

A completely different alignment concept is based on the determination of the charge neutrality level (CNL) or branch point (BP)
energy $E_{\rm BP}$ \cite{Bechstedt.Enderlein:1988:Book,Guo.Li.ea:2019:JPCC,Monch:2001:Book}. The use of such universal
reference level has been first suggested by Frensley and Kroemer \cite{Frensley.Kroemer:1976}. Apart from empirical tight-binding
descriptions of the BP energies \cite{Monch:2001:Book,Moench:1996:JoAP}, three different methods, which can be
combined with electronic structure calculations based on DFT and QP approaches, have been applied in the last decades: a Green function
method \cite{Tersoff:1984:PRB}, a determination through the zero of the integral of the density of states \cite{Guo.Li.ea:2019:JPCC,Robertson:2013:JoVSTA},
and the calculation from an approximate weighted sum of conduction- and valence-band energies \cite{Schleife.Fuchs.ea:2009:APL}. All these treatments
require only bulk calculations. Here we apply the third method, where the BP is computed as an average over the BZ and bands:
\cite{Schleife.Fuchs.ea:2009:APL}
\begin{equation}\label{eq2}
E_{\rm BP}=\frac{1}{2N}\sum_{\bf k}\left[\frac{1}{N_{\rm CB}}\sum^{N_{\rm CB}}_{c=1}\varepsilon_c({\bf k})+\frac{1}{N_{\rm VB}}
\sum^{N_{\rm VB}}_{v=1}\varepsilon_v({\bf k})\right] \,,
\end{equation}
where $N$ is the number of {\bf k}-points, and for a 2H unit cell the number of conduction bands is fixed to $N_{\rm CB}=2$ and that of valence bands to $N_{\rm VB}=4$.

Band alignment obtained using the branch point method of \eqref{eq2} are listed in Table~\ref{tab3} and displayed in Fig.~\ref{fig3}. Besides the hexagonal
Si$_x$Ge$_{1-x}$ alloys, results are also shown for 3C-Ge and 3C-Si for comparison.  For the cubic crystals we applied two different numerical 
descriptions of the electronic structure and BP energy. We studied the primitive diamond cell, with the resulting fcc BZ and halved number of
bands in \eqref{eq1}, as well as the same atomic arrangement described using a non-primitive hexagonal cell together with the corresponding hexagonal BZ.
The two procedures are indicated by 3C and 3C(hex), respectively. The first description is more appropriate for free standing diamond
crystals, while the second one, 3C(hex), better accounts the situation of band lineups at cubic [111]/ hexagonal [0001] interfaces.
We can conclude, first of all, that both approximate QP methods for band structures, namely MBJLDA and HSE06, can be applied with comparable results. Apart from the case
of 3C-Si, the deviations are in fact smaller than 0.1~eV,. This latter value can be used to define the uncertainty in the determination of the natural band
discontinuities $\Delta E_v$, $\Delta E_c$, and $\Delta E^{\rm ind}_c$ of heterostructures between systems with the smallest difference in composition (for the exact values see
Table~\ref{tab3}). For low (high) Si content the MBJLDA band edges are higher
(lower) than the HSE06 ones. Nevertheless, the results in Fig.~\ref{fig3} and Table~\ref{tab3} show that both approximate QP procedures can
be applied to the SiGe systems with qualitatively similar results. Since, however, minor numerical efforts are necessary to perform MBJLDA calculation, we will apply in the following only this approach to describe
the true band lineups.

\begin{sidewaystable}
\caption{Energies of the band extrema $E_c$ and $E_v$ at the $\Gamma$ point with respect to the BP position, taken as energy
zero as computed with MBJLDA and HSE06 functionals (values given in parenthesis). In addition, the conduction band
minimum at the $U$ point on the LM line, $E^{\rm ind}_c$, is listed, if the alloy is an indirect semiconductor. The natural band discontinuities
$\Delta E_c=E^{\rm lower}_c-E^{\rm upper}_c$ \eqref{eq1} and $\Delta E_v=E^{\rm upper}_v-E^{\rm lower}_v$, as well as the discontinuity for the indirect
CBM $\Delta E^{\rm ind}_c$ are also given. Because of the three possible configurations C1, C2, C3 and their average, four offsets are displayed for either
Si$_{0.25}$Ge$_{0.75}$/Si$_{0.5}$Ge$_{0.5}$ or Si$_{0.5}$Ge$_{0.5}$/Si$_{0.75}$Ge$_{0.25}$ heterostructures. The lowest and highest values describe
the 2H-Si/3C-Si and 3C-Ge/2H-Ge band lineups, respectively. Thereby, we distinguish between the band positions of 3C obtained using the diamond primitive cell with the fcc BZ for the BP determination (3C) or the equivalent hexagonal representation with the BP determination by integration over the smaller hexagonal BZ (denoted by 3C(hex)).
The uppermost (lowest) two lines describe band offsets between 3C (in one of the two versions) and 2H. All energy values are in eV.}

\begin{ruledtabular}
\begin{tabular}{ccccccccccl}
 
   alloy/polytype & $ E_{v}$ &  $ E_{c}$  & $ E_{c}^{\mathrm{ind}}$  &  $ \Delta E_{v}$ &   $ \Delta E_{c}$ &   $ \Delta E_{c}^{\mathrm{ind}}$  \\ 
 \hline

 \
3C-Ge           &0.221(0.300) &0.929(0.943) &0.877(0.935)$^{\mathrm{L}}$ &  $-$0.039(0.095) &$-$0.363($-$0.443) &  0.011($-$0.115) \\
3C-Ge(hex)       & 0.027(0.061) &0.683(0.692)   &0.652(0.645)$^{\mathrm{L}}$ &$-$0.233($-$0.144)  & $-$0.117($-$0.192) & 0.236(0.175)     \\      
2H-Ge  &0.260(0.205) &0.566(0.500) &0.888(0.820)$^{\mathrm{U}}$ &$-$0.006($-$0.005)  &  0.230(0.263)  &  0.158(0.143)\\
Si$_{0.25}$Ge$_{0.75}$ &0.266(0.200) &0.796(0.763) &1.046(0.963)$^{\mathrm{U}}$ &0.183(0.002) &  0.323(0.450)  & $-$0.048(0.085)\\
&&&&0.146($-$0.042)  &  0.385(0.520) &  $-$0.045(0.104)\\

Si$_{0.50}$Ge$_{0.50}\mathrm{(C1)}$ &0.083(0.198) &1.119(1.213) &0.998(1.048)$^{\mathrm{M}}$ &$-$0.03($-$0.148) &  0.142(0.233)  & $-$0.095(0.042)\\
Si$_{0.50}$Ge$_{0.50}\mathrm{(C2)}$ &0.120(0.242) &1.181(1.283) &1.001(1.067)$^{\mathrm{U}}$& 0.100($-$0.045)  &  0.283(0.456)  & $-$0.063(0.077)\\
Si$_{0.50}$Ge$_{0.50}\mathrm{(C3)}$ &0.296(0.348) &0.938(0.996) &0.951(1.005)$^{\mathrm{M}}$ &0.029(0.008)   &  0.235(0.236)  & $-$0.036($-$0.036)\\
Si$_{0.50}$Ge$_{0.50}\mathrm{(average)}$ &0.166(0.262) &1.079(1.164)&0.983(1.040) &0.066(0.052)	  &  0.173(0.166)  & $-$0.039($-$0.055)\\
&&&&0.242(0.158)	  &  0.416(0.453)  &  0.011(0.007)\\
&&&&0.112(0.072)	  &  0.275(0.285)  & $-$0.028($-$0.021)\\
Si$_{0.75}$Ge$_{0.25}$ &0.054(0.190) &1.354(1.449) &0.962(1.012)$^{\mathrm{M}}$& 0.165(0.190) &  0.246(0.238)  & $-$0.045($-$0.058)\\
2H-Si &\llap{$-$}0.111(0.000) &1.600(1.687) &0.917(0.954)$^{\mathrm{M}}$ &0.241(0.307)  &  0.298(0.243)  & $-$0.019($-$0.124)\\
3C-Si(hex)     & $-$0.352($-$0.307)&1.898(1.930)& 0.898(0.830)$^{\mathrm{M}}$&0.155(0.115) &  $-$1.313( $-$1.493)  &  $-$0.067($-$0.061) \\
3C-Si &\llap{$-$}0.266($-$0.115) &2.913(3.180) &0.984(1.015)$^{\mathrm{X}}$ 
     
\label{tab3}
\end{tabular}
\end{ruledtabular}
\end{sidewaystable}

From a qualitative analysis of Fig.~\ref{fig3} and Table~\ref{tab3} we can identify a clear tendency for a type-I heterostructure
\cite{Bechstedt.Enderlein:1988:Book,Kittel:2005:Book} when hexagonal Si$_x$Ge$_{1-x}$ and Si$_y$Ge$_{1-y}$ alloys with $x<y$ are combined.
Quantum wells are formed in the layer with the lower Si content for both electrons and holes. However, for Ge-rich alloys the heterostructure type is not very clear. The position of the top of the valence band is weakly dependent on the composition. In general, see Table~\ref{tab3}, the absolute values of the valence band offsets are smaller than the conduction band ones. Our results agree qualitatively with the predictions of Wang et al.\cite{Wang.Zhang.ea:2021:APL}. More precisely, we can compare the minimum fundamental gaps of
2H-Si$_x$Ge$_{1-x}$, calculated using MBJLDA (HSE06), are $E_g$ or $E^{\rm ind}_g$ = 0.31 (0.30), 0.53 (0.56), 0.82 (0.78), 0.91 (0.82) and
1.03 (0.95)~eV from Table~\ref{tab3} and the values $E_g$ or $E^{\rm ind}_g = 0.30$, 0.65, 0.90, 0.94, and 1.06~eV from Ref.\onlinecite{Wang.Zhang.ea:2021:APL}.
The trends on the band edge position are also very similar. Only the fact, visible in Fig.~\ref{fig3}, that $E_v$ is above the BP energy for the Si$_x$Ge$_{1-x}$ alloys (apart from the 2H-Si case) is less pronounced in Ref.~\cite{Wang.Zhang.ea:2021:APL}.

Interesting byproducts of the band lineups in Fig.~\ref{fig3} and Table~\ref{tab3} are the band offsets between the 3C and 2H
polytypes of Ge and Si. We can see that the description of the cubic polytypes using the diamond cell, 3C, or the hexagonal representation strongly influences 
the resulting natural band offsets. Low conduction-band states at 2/3 $\Gamma$L (diamond) are mapped onto the $\Gamma$ (hexagonal representation) point.
For Si a drastic lowering of the direct CBM in 3C(hex) occurs by about 1eV compared to the 3C case. This can be understood by the simple band-folding arguments 
that the second CBM at ${\Gamma}$ of 3C-Si(hex) at 3.179 eV almost agrees in energy with the first CBM at ${\Gamma}$ in 3C-Si, see Table~\ref{tab3}. 
It is also worth to mention that the CBM near M in 3C-Si(hex), located at 2.15 eV, agrees with the position of the indirect CBM near ${\mathrm{X}}$ in 3C-Si.
In Ge the energy positions vary on an absolute scale because of the resulting different BPs. The BP in 3C-Ge(hex) is by about 0.2 eV
higher in energy compared to the BP of 3C-Ge, but still below the VBM, independently of the chosen QP approximation. In close agreement to the vacuum level alignment, 
the BP alignment starting from MBJLDA and HSE06 electronic structures leads to similar heterostructure characters for 2C/2H junctions. However, the absolute values, sometimes
also the signs, of the band discontinuities depend on the choice of the structure, 3C(hex) or 3C, and the direct or indirect nature of the gap. We focus the discussion on the heterocharacter of 2H/3C on the lowest conduction bands at $\Gamma$  (outside $\Gamma$) for Ge (Si) and identify the conduction band discontinuities by $\Delta E_{c}$  ($ \Delta E_{c}^{\mathrm{ind}}$). 

Using MBJLDA (HSE06), as we can deduce from Table~\ref{tab3}, delivers $ \Delta E_{c} = -0.36 (-0.44)$ or $-$0.12 ($-$0.19) and $ \Delta E_{v}$ = $-$0.04 (0.10) or $-$0.23 ($-$0.14) eV for Ge 2H/3C
or 2H/3C(hex) heterostructures, while the corresponding values for Si heterostructures are $\Delta E_{c}$ = $-$0.02 ($-$0.12) or $-$0.07 ($-$0.06) eV and $\Delta E_{v}$ = 0.24 (0.31) or 0.16 (0.12) eV for 2H/3C(hex) or 2H/3C. Consequently, the heterojunction character varies somewhat with the selected QP approximation and the choice of the unit cell for the calculation of the approximate BP energy. Within the MBJLDA approximation, the heterocharacter tends generally 
 toward type-I. Only the Si-based 2H/3C(hex) junction exhibits a type-II character. These findings are qualitatively similar to the results using the vacuum-level aligment shown in Fig.~\ref{fig2}. Using the HSE06 functional for the determination of the band structures, the 2H/3C(hex) Ge junction has a type-I lineup, while a type-II interface results for the Si case. These findings are in agreement with 
 other HSE06 calculations for 'true' band offsets using supercells \cite{Kaewmaraya-jpcc-2017}.  \par
 
 The uncertainties in the prediction of the type of 3H/3C or 2H/3C(hex) interfaces using the natural band lineup remember similar findings for InP and
  GaAs \cite{Bechstedt.Belabbes:2013:JoPCM}. Due to fluctuations in the bond stacking during growth of GaAs quantum wires indeed such quantum 
  wells appear, where, however, the heterostructure type~I and II is under discussion \cite{Spirkoska.Arbiol.ea:2009:PRB,Akopian.Patriarche.ea:2010:NL}.
  The same statement is valid for other calculations of the heterocrystalline interface 2H/3C of silicon using several approaches. The band alignment method
  tends to type-II band lineup \cite{Wang.Zhang.ea:2021:APL,Raffy-PRB-2004,Amato.nanolett-2016}, whereas hexagonal/cubic Si superlattice (see also below)
  show type-I band offsets \cite{Raffy.Furthmueller.ea:2002:PRB,Murayama.Nakayama:1994:PRB,Kaewmaraya-jpcc-2017}. In the case of Ge the situation is more
   difficult, because of the metallic character of the material in DFT \cite{Raffy.Furthmueller.ea:2002:PRB}. Moreover, the band alignment method (type-II) 
   \cite{Wang.Zhang.ea:2021:APL} and the superlattice method (type-I) \cite{Kaewmaraya-jpcc-2017} again suggest opposite heterostructure characters.

 For the BP position using the VBM as a energy zero we find $E_{\rm BP}=-0.22$ ($-$0.30), $-$0.03 ($-$0.06), $-$0.26 ($-$0.21), 0.11 (0.00), 0.35 (0.31), and 0.27 (0.12)~eV for 3C-Ge,
 3C-Ge(hex), 2H-Ge, 2H-Si, 3C-Si(hex) and 3C-Si, respectively, using MBJLDA (HSE06). In the case of 3C-Si, there is a good agreement with other calculations of $E_{\rm BP}=0.29$
\cite{Hoffling.Schleife.ea:2012:PRB}, 0.16 \cite{Hinuma.Grueneis.ea:2014:PRB}, 0.03 \cite{Monch:2001:Book}, 0.36 \cite{Tersoff:1984:PRB},
and 0.2~eV \cite{Robertson:2013:JoVSTA}. The BP positions in 3C-Ge are $E_{\rm BP}=-0.28$ \cite{Hinuma.Grueneis.ea:2014:PRB},
$-$0.28 \cite{Monch:2001:Book}, 0.18 \cite{Tersoff:1984:PRB}, and 0.1~eV \cite{Robertson:2013:JoVSTA}. 
Despite similarities with the gap results of Wang et al. \cite{Wang.Zhang.ea:2021:APL}, a clear discrepancy appears for the positions of the CNL with respect to the
VBM for both 3C- and 2H-Ge. In the latter references SOC has not been included. The tendency predicted here to find negative BP values for Ge is confirmed both by
\emph{ab initio} GW as well as empirical tight-binding calculations. Such a prediction of a BP energy in a band may have significant consequences for the occurrence of $p$-type
accumulation at Ge surfaces, as demonstrated for the surface electron accumulation of In compounds, e.g. In$_2$O$_3$
\cite{King.Veal.ea:2009:PRBa}. The BP positions allow to define a natural valence band discontinuity $\Delta E_v$ between
3C-Ge and 3C-Si or 3C-Ge(hex) and 3C-Si(hex) of $\Delta E_v=0.49$ (0.42)~eV or 0.38 (0.37)~eV (see Fig.~\ref{fig3}). These values are close to that $\Delta E_v=0.38$~eV of GW
calculations \cite{Hinuma.Grueneis.ea:2014:PRB} and that $\Delta E_v=0.34$~eV measured by photoemission \cite{Katnani.Margaritondo:1983:PRB}.

\subsection{\label{sec3c} Comparison of alignment methods}

Tables~\ref{tab2} and \ref{tab3}, as well as Figs.~\ref{fig2}(b) and \ref{fig3}, show the natural band discontinuities $\Delta E_{v}$, $ \Delta E_{c}$ and $\Delta E_{c}^{\mathrm{ind}}$ of
MBJLDA band structures, derived with different alignment procedures, namely the vacuum level alignment and the BP alignment. The resulting heterojunction behaviors agree qualitatively, but exhibit quantitative differences. The smallest deviations occur for the offsets of the VBM. For Ge(Si)-rich alloy heterostructures the two $ \Delta E_{v}$ values differ by $-$9 meV (23 meV). The differences
increase toward 88 meV ($-$81 meV) including alloys with the same numbers of Si and Ge atoms in the heterojunction. The discrepancies are similar for the CBM  offsets and increase for the CBM$^{\mathrm{ind}}$ offsets.
For more Ge-rich heterostructures the direct CBM offsets vary by $-$19 or $-$52 meV, while for Si-rich junctions the indirect CBM offsets deviate by 81 or $-$24 meV. All these deviations are smaller than the accuracy
of the underlying approximate QP methods for the gaps, that we estimate to be smaller than 0.1eV \cite{Bechstedt:2015:Book}.

Interestingly, both alignment procedures also nearly give the same natural band lineups for the two heteropolytypic structures 3C/2H (see Figs.~\ref{fig2}(b) and \ref{fig3}, as well as Tables~\ref{tab2} and \ref{tab3}). Results of both 
alignments tend toward a type-I heterocharacter with deep electron and flat hole quantum wells in the region of the 2H polytype. In the Si case the agreement is complete for $\Delta E_{v}$ and $ \Delta E_{c}$. Only
the positions of the indirect CBM differ by 81 meV. In the Ge case small deviations of only 39 meV ($\Delta E_{v}$), 31 meV ($\Delta E_{c}$) and 31 meV ($\Delta E_{c}^{\mathrm{ind}}$) are visible. The minor deviations of the band discontinuities between two polytypes 3C and 2H, as well as between two hexagonal alloys suggest the applicability of both alignment procedures. 
Nevertheless, we focus in the following on the BP alignment because it relies only on bulk calculations.

\section{\label{sec4}True band lineup}
\subsection{\label{sec4a}Interfacial biaxial strain}

A heterostructure 2H-Si$_x$Ge$_{1-x}$/2H-Si$_y$Ge$_{1-y}$ ($x<y$) may be fabricated by controlled growth in the direction of the $c$-axis. For not too large
differences in $x$ and $y$ lattice parameters, pseudomorphic epitaxial growth should be possible. Neglecting the interface mixing of regions with different compositions, the main effect of the pseudomorphic growth will be biaxial strain at the junction, resulting in
compressive strain of the alloy with lower Si content and tensile strain of the Si-richer mixed crystal. Since the maximum lattice mismatch
$f=2\frac{a(x)-a(y)}{a(x)+a(y)}$ amounts to 4.3~\% (see Table~\ref{tab1}), the resulting mismatch of Si$_n$Ge$_{4-n}$/Si$_{n+1}$Ge$_{3-n}$
($n=0,1,2,3$) interfaces is approximately 1~\%, i.e., still in the validity range of Hooke's law. 
Setting the in-plane lattice constant  of the heterojunction equal to $a$, assuming pseudomorphic growth, the resulting biaxial strain on the two sides of the heterointerface 
Si$_x$Ge$_{1-x}$/Si$_{y}$Ge$_{1-y}$
amounts to
\begin{equation}\label{eq3}
\epsilon_\|(x/y)=\frac{a-a(x/y)}{a(x/y)}
\end{equation}
where  $a(x)$ and $a(y)$ are the lattice constant of the two bulk alloys.
The band energies are shifted from the values $E_v$, $E_c$ and $E^{\rm ind}_c$ for the unstrained alloys to the values
$E_v(\epsilon_\|)$, $E_c(\epsilon_\|)$, and $E^{\rm ind}_c(\epsilon_\|)$. In the framework of the validity of the Hooke's law
\cite{Kittel:2005:Book} a linear strain dependence holds ($\nu=v,c,c^{\rm ind}$):
\begin{equation}\label{eq4}
E_\nu(\epsilon_\|)=E_\nu+\Xi_\nu\epsilon_\|.
\end{equation}
The band deformation potentials listed in Table~\ref{tab4} have been computed from the MBJLDA band energies of the unstrained and strained
alloys given in the same table. Thereby, an arithmetic average of the deformation potentials arising for $\epsilon_\|=\pm0.01$ is used,
in order to avoid nonlinear effects beyond the Hooke's law. Since the band energies are computed with respect to the BP energy \eqref{eq2}, the resulting
deformation potentials have to be interpreted as the deformation potentials of the energy difference between band extrema and
BP energy. \par
The general effects of a biaxial strain on the band edges, measured with respect to the BP energy zero, are illustrated in Fig.~\ref{fig4}.
In the limit of compressive strain, the $a$-lattice constant is shortened and the band gaps at $\Gamma$ are increased, although, in addition to the CBM, $E_c$, also the VBM, $E_v$, is shifted away
from the BP. In the case of tensile strain the opposite shifts are visible, in particular the gaps get shrunk. Interestingly, the conduction band minima
along the LM line hardly vary with strain, at least with respect to the BP used for band alignment. Moreover, the character of the heterostructure (type I or II) is rather
independent of the biaxial strain. The type-I character of the 2H/3C heterostructures (with MBJLDA) with 2H as quantum-well and 3C as barrier material, known from
the natural band lineup, is conserved. The indirect character of the band gap is more pronounced under compressive strain, but disappears for tensile strain.

\begin{table*}[h!]
\caption{Energies of band extrema $E_v$, $E_c$ and $E^{\rm ind}_c$ measured to the BP position of SiGe alloys and polytypes under
biaxial strain $-$1, 0, +1~\%. The MBJLDA approach is applied. The deformation potentials resulting from the strain-induced band
displacements are also given. All values are in eV.}
\centering
\footnotesize
\begin{ruledtabular} 
\begin{tabular}{ccccccccccccccccl}
Material   & \multicolumn{3}{c}{Compressive} & \multicolumn{3}{c}{Zero} & \multicolumn{3}{c}{Tensile} & \multicolumn{3}{c}{ Resulting }  \\
  & \multicolumn{3}{c}{biaxial strain $-1\%$}&\multicolumn{3}{c}{biaxial strain $0\%$}&\multicolumn{3}{c}{biaxial strain $1\%$ } &\multicolumn{3}{c}{deformation potential } \\ 
  \cline{2-4} \cline{5-7}  \cline{8-10} \cline{10-13} \\
   Si$_{x}$Ge$_{1-x}$& $E_{v}$ &   $ E_{c}$   &    $ E_{c}^{\mathrm{ind}}$& $E_{v}$ &   $ E_{c}$   &    $ E_{c}^{\mathrm{ind}}$   &  $E_{v}$ &   $ E_{c}$   &    $ E_{c}^{\mathrm{ind}}$ & $\Xi_{v}$ & $\Xi_{c}$  & $\Xi_{c}^{\mathrm{ind}}$ \\ 
 \hline

3C-Ge                               & 0.248 & 1.197 & 1.031 &        0.221  & 0.929  & 0.877 &   0.195 & 0.656 & 0.715                                                    & $-$2.65 & $-$27.15 &$-$15.80  \\
3C-Ge(hex)                         &0.05 &0.834&0.676&           0.027  &0.683    &0.652&    0.004&0.525&0.628 &                                                     $-$2.30&$-$15.45&$-$2.36       \\               	
2H-Ge                               & 0.321 & 0.788 & 0.876 &    0.260 & 0.566 & 0.888 &  0.213 & 0.364 & 0.897                                                        & $-$5.40& $-$21.15 &1.07 \\
Si$_{0.25}$Ge$_{0.75}$   & 0.334 & 1.003 & 1.088 &  0.266  & 0.796 & 1.046  &  0.223 & 0.647 & 1.003                                                        &  $-$5.55 & $-$17.83 & $-$4.25 \\
Si$_{0.50}$Ge$_{0.50}(\mathrm{C1}$)    & 0.132 & 1.354 & 1.045 &  0.083 &1.119 & 0.998 &                0.034 & 0.899 & 0.949                       & $-$4.90& $-$22.75 &$-$4.80\\
Si$_{0.50}$Ge$_{0.50}(\mathrm{C2}$)   & 0.173 & 1.412 & 1.046 &  0.120	&1.181  & 1.001&                  0.067 & 0.965 & 0.959                       & $-$5.30 & $-$22.40 & $-$4.30\\
Si$_{0.50}$Ge$_{0.50}(\mathrm{C3}$)   & 0.360 & 1.124 & 1.012 &  0.296 & 0.938 & 0.951 &                0.231 & 0.744& 0.895                        & $-$6.45 &$-$19.00 & $-$5.85 \\
Si$_{0.50}$Ge$_{0.50}\mathrm{(average)}$   & 0.221 & 1.296 & 1.034 &  0.166  & 1.079 & 0.983   &  0.110 & 0.869 & 0.934                       & $-$5.55 &$-$21.38 &$-$5.00 \\
Si$_{0.75}$Ge$_{0.25}$    & 0.126 & 1.573 & 1.023 &  0.054  & 1.354 & 0.962 & $-$0.005 & 1.139 & 0.906                                            & $-$6.55 & $-$21.68 &$-$5.95 \\
2H-Si                                     & \llap{$-$}0.043 & 1.827 &  0.975 & $-$0.111  & 1.600 & 0.917  & $-$0.170 & 1.372 & 0.859            & $-$6.35 &$-$22.75 & $-$5.80 \\
3C-Si(hex)                                  & \llap{$-$}0.346 & 2.116 & 0.906 & $-$0.352&1.898& 0.898  & $-$0.360 &1.673 & 0.910                 & $-$0.68 & $-$22.13& 0.17 \\
3C-Si                                     & \llap{$-$}0.236& 2.945 & 0.951 & $-$0.266 & 2.913 & 0.984  & $-$0.293 & 2.880 & 1.000              & $-$2.85 &$-$3.30 & 2.45
\label{tab4}
\end{tabular}
\end{ruledtabular}
\end{table*}

To predict `true' band discontinuities by estimating the biaxial strain at Si$_x$Ge$_{1-x}$/Si$_y$Ge$_{1-y}$ interfaces, we use the band deformation potentials in Table~\ref{tab4}. However, they are calculated with respect to the BP, which cannot be measured directly.  For comparison with measurements it is better to study the deformation potentials of the gaps, $E_g=E_c-E_v$ and 
$E_{g}^{\rm ind}=E^{\rm ind}_c-E_v$.
In the cubic limit biaxial deformation potentials $\Xi=-13.15$ ($-$21.45)~eV and $\Xi^{\rm ind}=0.06$ (0.85)~eV result for 3C-Ge (3C-Si) but using
the hexagonal BZ to describe the high-symmetry points $\Gamma$, L, and M. Our estimates of deformation potentials of 3C-Ge at the
$\Gamma$ and L points of the fcc BZ deliver $\Xi(\Gamma)=-24.5$~eV and $\Xi(\mathrm{L})=-13.15$~eV. Because of the projection of the L point in the fcc BZ
onto the $\Gamma$ point of the hexagonal BZ, the indirect gap value in the cubic case $\Xi(\mathrm{L})$ should be comparable with the $\Xi=-13.15$ eV value
estimated for the direct gap of 3C-Ge, using a hexagonal {\bf k}-point sampling. Independently of the description, in diamond-Si the indirect gaps possess
positive deformation potentials of 5.30 eV for 3C-Si and 0.85 eV for 3C-Si(hex). Our estimates approach measured deformation potentials at the zone boundaries
\cite{Yu.Cardona:1999:Book}. In the hexagonal alloys the deformation potentials $\Xi$ for the direct gaps relatively weakly vary with the composition
around $\Xi= -21$ eV. Only for $x=0.25$ a slightly smaller absolute value with $\Xi=-17.83$~eV is estimated. The situation is completely different for the deformation
potential of the lowest indirect gap. Because of similar values of the CBM with mixed $sp$ character and the $p$-like VBM the gap deformation potentials remain small.
An exception appears for 2H-Ge, for which the indirect gap deformation potential tends to be small of 6.5 eV but positive.

\subsection{\label{sec4b}Interface influence}

The band lineups of conduction and valence bands can be simulated for supercells containing an itterface between the two materials, forming a heterostructure. In general, the
theoretical modeling of interfaces is difficult because of possible lattice mismatches, crystal structure misfits, heterovalencies, chemical and structural
disorder, and presence of defect states. In the case of hexagonal Si$_x$Ge$_{1-x}$ alloys, grown pseudomorphically on top of each other, one can focus on
the lattice mismatch and the configuration of Si and Ge atoms on the atomic sites at the heterointerface. A powerful tool is the superlattice
method with superlattice structures based on hexagonal Si$_x$Ge$_{1-x}$/Si$_y$Ge$_{1-y}$ heterostructures in the [0001] direction building a superlattice unit cell.

\begin{figure}[h]
\includegraphics[width=12cm]{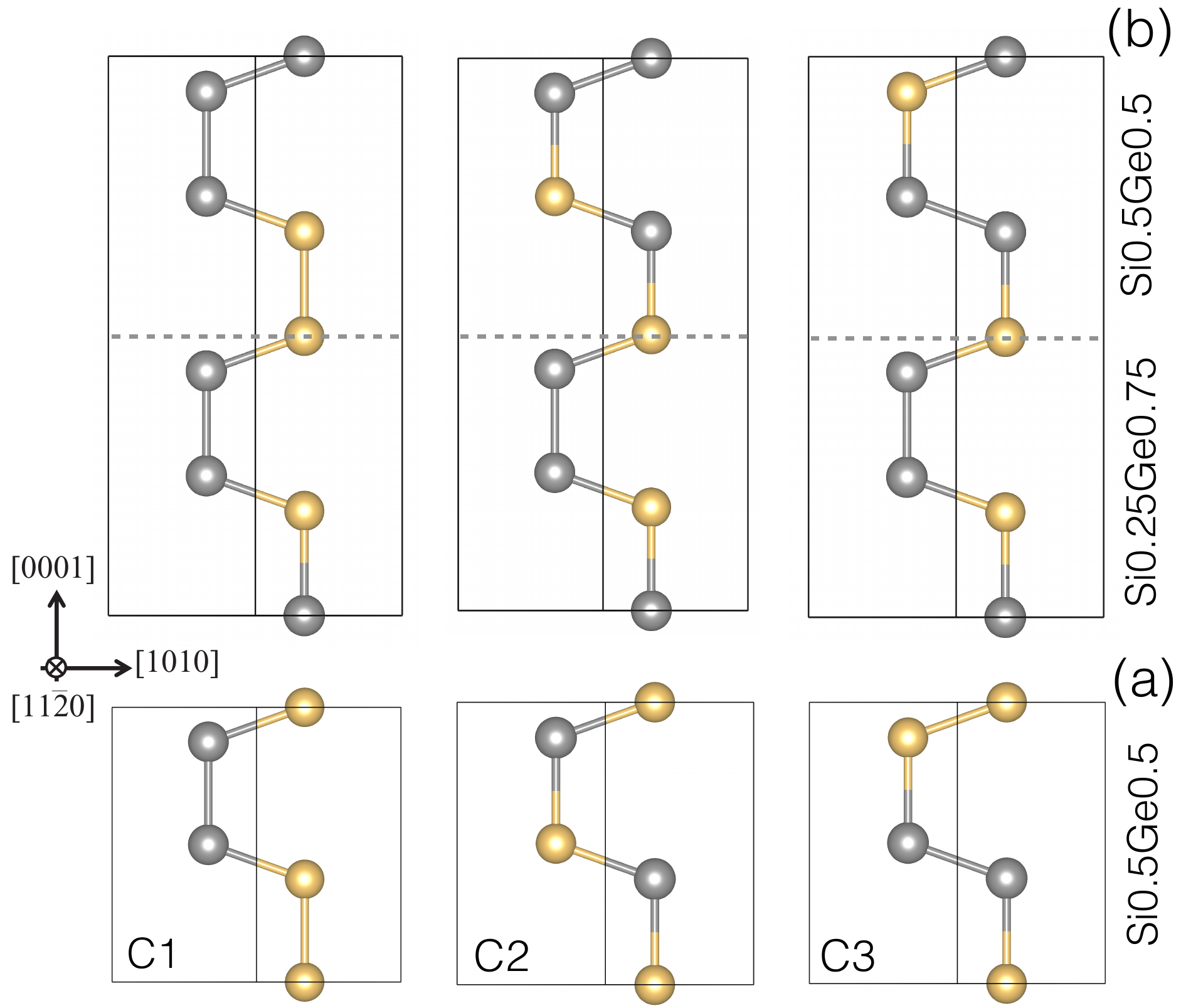} 
\caption{(a) The side view is shown for three possible symmetry-inequivalent bond stackings and atomic configurations C1, C2 and C3 of the hexagonal Si$_2$Ge$_2$
alloy. (b) The crystal structure of three different atomic arrangements of hexagonal Si$_2$Ge$_2$  alloys (C1, C2, and C3 configurations) as heterointerface with Si$_1$Ge$_3$
to simulate Ge-rich junction. The yellow and gray balls represent Si and Ge atoms, respectively. The nominal interfaces are indicated
 by dashed horizontal lines between two unit cells stacked in $c$-axis direction.}
\label{fig1}
\end{figure}

\begin{figure}[h]
\includegraphics[width=17cm]{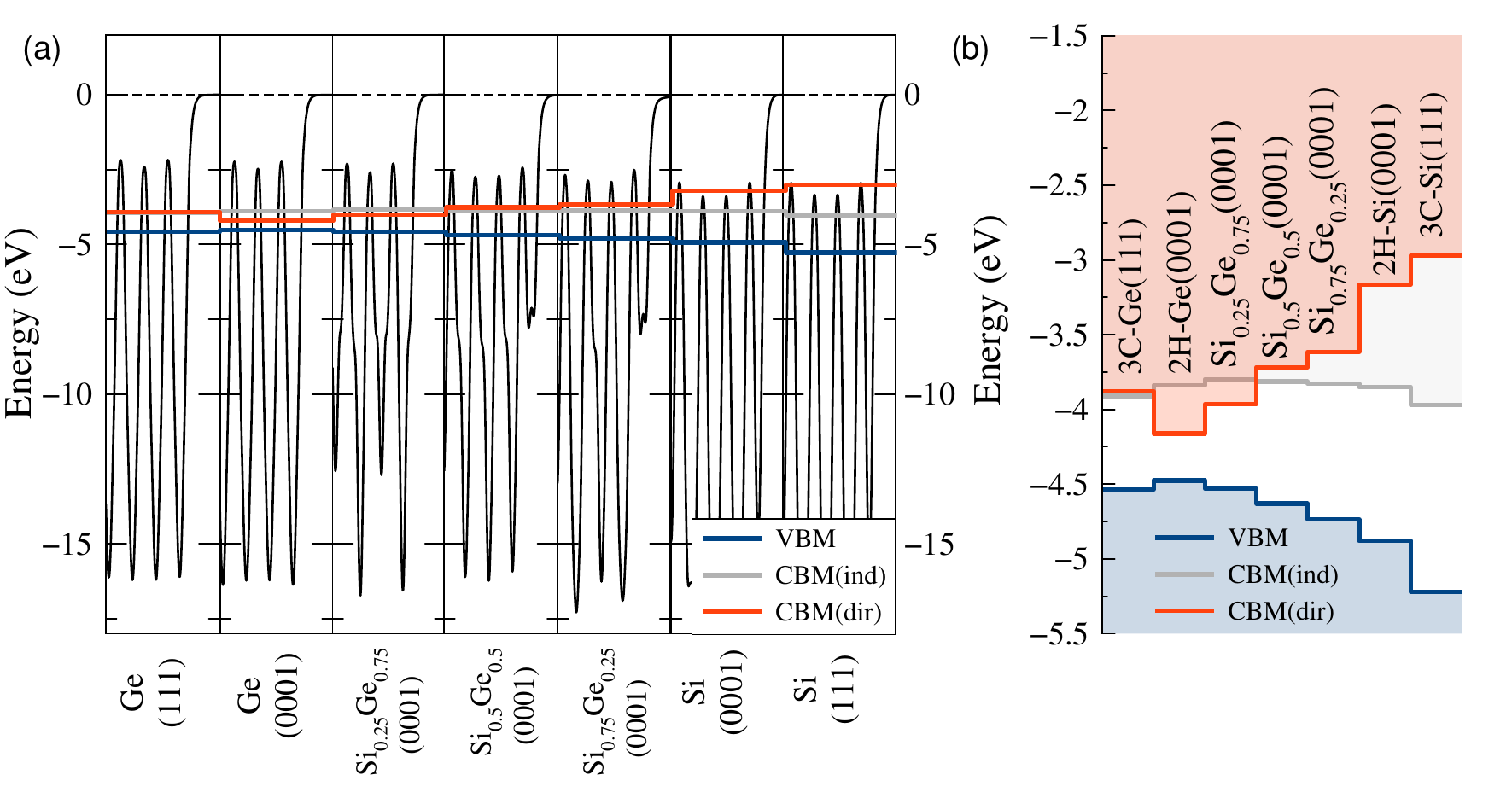} 
\caption{Alignment procedure and resulting band lineup for heterostructures consisting of hexagonal Si$_{x}$Ge$_{1-x}$ alloys.
For comparison also results for the 3C polytypes oriented in [111] direction are given. The MBJLDA framework is used for the electronic structure calculations.
 (a) Averaged electrostatic potentials (black curves) plotted along the $c$-axis, i.e., the [0001] direction. The positions of the bulk band extrema 
 $E_c$ (in red),  $E^{\rm ind}_c$ (in grey) and $E_v$ (in blue) are also given. (b) The band lineups resulting with the electron affinity rule. The vacuum level is taken as energy zero.}
\label{fig2}
\end{figure}

\begin{figure}[h]
\includegraphics[width=14cm]{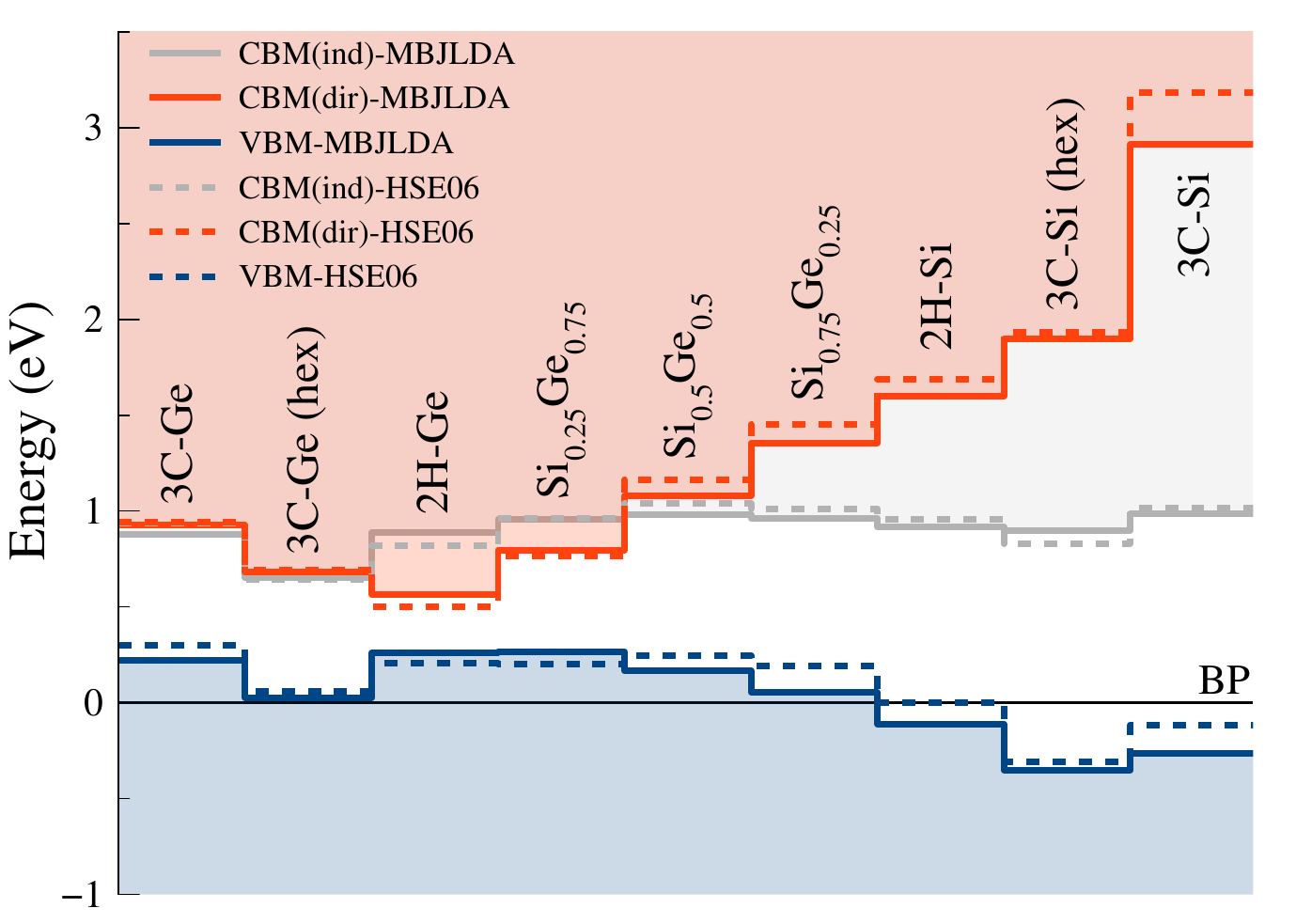} 
\caption{Band lineups applying unstrained bulk positions of the band extrema, the VBM at $\Gamma$ ($E_v$), the CBM at $\Gamma$ ($E_c$),
and the CBM at the LM line ($E^{\rm ind}_c$). The conduction band edges are displayed in red ($E_c$) or yellow ($E^{\rm ind}_c$), while the
top of the valence bands is shown in blue. The black horizontal line defines the BP, the energy zero. Solid (dashed) lines are computed in the
MBJLDA (HSE06) framwork. In the Si$_{0.5}$Ge$_{0.5}$ case only averaged energies (see Table~III) are plotted.}
\label{fig3}
\end{figure}

\begin{figure}[h]
\includegraphics[width=16cm]{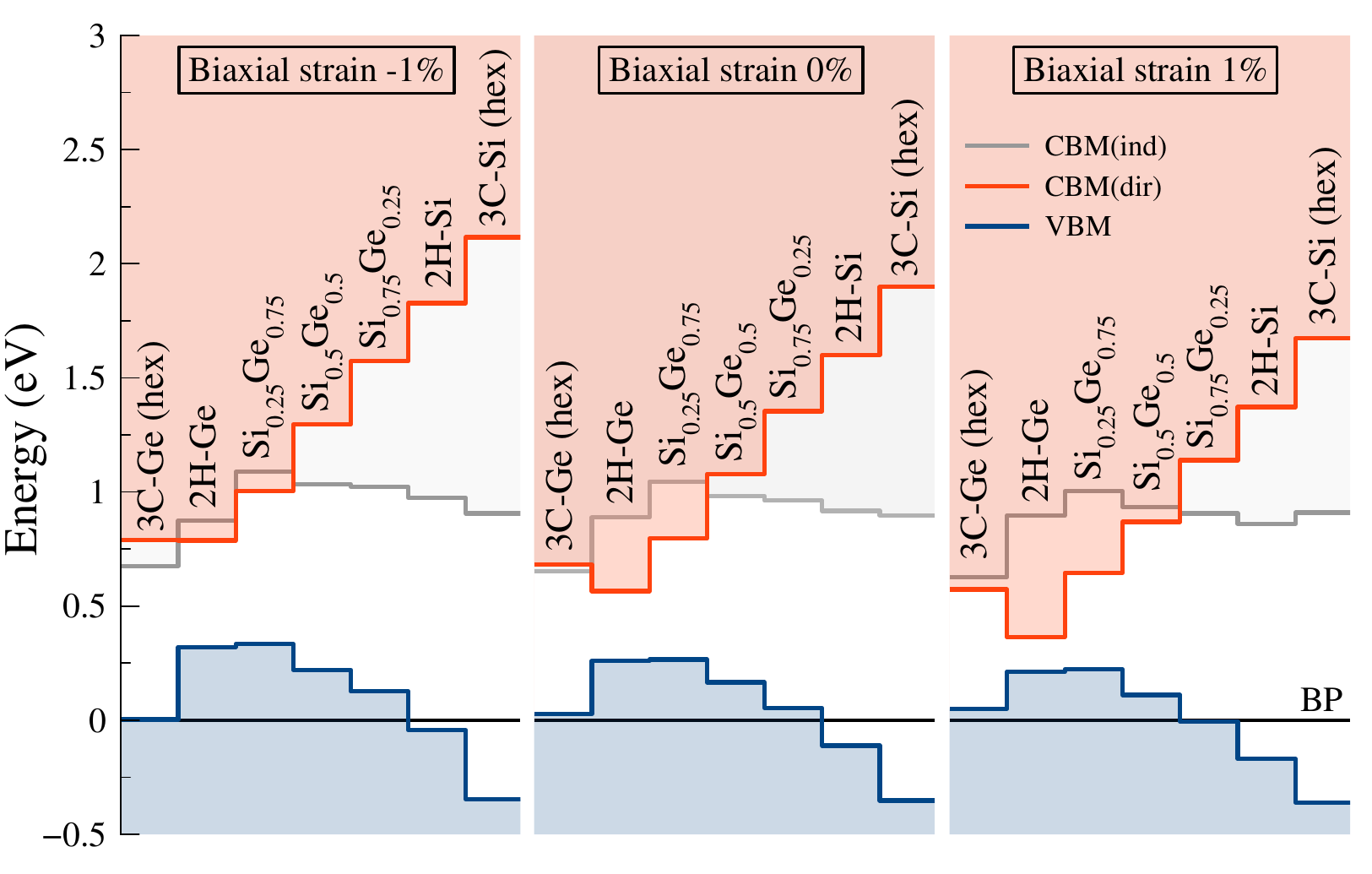} 
\caption{Band lineups of Si$_x$Ge$_{1-x}$ alloys and polytypes under compressive, zero and tensile biaxial strain, $\epsilon_\|=-0.01$, 0.00, and 0.01,
drawn with respect to the BP as energy zero. $E_c$: red, $E^{\rm ind}_c$: grey, and $E_v$: blue. In the Si$_{0.5}$Ge$_{0.5}$ case energies
averaged over the three configurations shown in Fig.~\ref{fig1} are displayed.}
\label{fig4}
\end{figure}

In the simplest case we can match two hexagonal unit cells of Si$_n$Ge$_{4-n}$ and Si$_{n+1}$Ge$_{3-n}$ ($n=0,1,2,3$), joining the surfaces perpendicular to the $c$-axis to create a hexagonal supercell of (Si$_n$Ge$_{4-n}$)$_{1}$ (Si$_{n+1}$Ge$_{3-n}$)$_{1}$(0001) with an in-plane lattice constant $a$ and a $c$-lattice constant that is approximately twice the value of the corresponding lattice constant of the isolated hexagonal alloys. The resulting superlattice
represents a hexagonal crystal with eight atoms, ($2n+1$) Si and ($7-2n$) Ge atoms in the unit cell with the lattice parameters $a$ and $c$. The highest
complexity of the atomic arrangements in such a superlattice cell appears if a stoichiometric alloy Si$_{2}$Ge$_{2}$ is included. Because of the existence of the
three configurations C1, C2, and C3 (see Fig.~\ref{fig1}(a)) several arrangements are possible at the interface. We consider three different
atomic arrangements at the heterointerface of Si$_{2}$Ge$_{2}$ with Si$_{1}$Ge$_{3}$ to simulate Ge-rich junctions, as shown in Fig.~\ref{fig1}(b). We start with
a minimum Si-Si bond distance along the $c$-axis, e.g. configuration C1, and then we consecutively move out one of the Si atoms in order to generate a maximum
of Si-Si atomic separations, e.g. configuration C3 from Fig.~\ref{fig1}(b). Although the difference between these atomic configurations is very small in terms of the
resulting lattice parameters, the actual atomic arrangement has a strong impact on the band-gap energy and the position of the CBM in \textbf{k}-space. A direct-indirect
gap crossover occurs in the set of (Si$_{1}$Ge$_{3}$)$_{1}$(Si$_{2}$Ge$_{2}$)$_{1}$(0001) superlattices. A similar procedure is applied for the  (Si$_{2}$Ge$_{2}$)$_{1}$(Si$_{3}$Ge$_{1}$)$_{1}$(0001) superlattices. For the two other superlattices  (Ge$_{4}$)$_{1}$(Si$_{1}$Ge$_{3}$)$_{1}$(0001) and  (Si$_{3}$Ge$_{1}$)(Si$_{4}$)$_{1}$(0001)
only one configuration has to be studied. All these short-period superlattice geometries have to be optimized by atomic relaxations and minimization of the total energy.
The resulting lattice parameters $a$ and $c$ are listed in Table~\ref{tab5}. In the cases of presence of Si$_{2}$Ge$_{2}$ alloys we only list the average values. 
These $a$-parameters, together with the $a$-lattice constants in Table~\ref{tab1}, allow to define the biaxial strain $\epsilon_\|$ \eqref{eq3} at the interface between two alloys.

\begin{table*}[h!]
\caption{Results of the superlattice approach and the macroscopic treatment using elastic moduli for four different heterostructures. Lattice
constants $a$ and $c$ (in {\AA}), biaxial strain $\epsilon_\|$ (in \%), and band energies $E_v$, $E_c$, and $E^{\rm ind}_c$ (in eV) estimated with
the deformation potentials in Table~\ref{tab4}. For the Si$_2$Ge$_2$ alloys energies are computed for the configurations C1, C2, C3 and their
average. Therefore, four different sets of band energies appear for Si$_{1}$Ge$_{3}$/Si$_{2}$Ge$_{2}$ and  Si$_{2}$Ge$_{2}$/Si$_{3}$Ge$_{1}$ heterostructures.
}

\begin{tabular}{cccccccccccc} \hline
 hetero-             & \multicolumn{3}{c}{super. app.} & \multicolumn{2}{c}{macro. app.} & \multicolumn{3}{c}{super. app.} & \multicolumn{3}{c}{macro. app.} \\ \cline{2-12}
structure           & $a$   & $c$ & $\epsilon_\|$ & $a$ & $\epsilon_\|$ & $E_v$ & $E_{c}$ & $E^{\rm ind}_c$ & $E_v$ & $E_{c}$ & $E^{\rm ind}_c$  \\ \hline
Ge$_4$             &              &                 & $-$0.96  &               & \llap{$-$}0.65  & 0.319  & 0.763  & 0.875     & 0.300 & 0.710 & 0.880 \\
Ge$_4$/Si$_1$Ge$_3$                        & 3.9577  & 13.0676  &                & 3.969  &                &             &            &            &            &            &    \\
Si$_1$Ge$_3$ &              &                 & {   }  0.29      &                 & 0.60         & 0.253  & 0.777  & 1.033   & 0.240 & 0.720 & 1.020 \\
Si$_1$Ge$_3$ &              &                  & $-$0.77 &               & \llap{$-$}0.61   & 0.316  & 0.944  & 1.078 & 0.307    & 0.918  &1.019  \\
Si$_1$Ge$_3$/Si$_2$Ge$_2$                       & 3.9161 & 12.9289   &           &   3.922            &                 &            &          &                &    &            &  \\
Si$_2$Ge$_2$ &               &                 & {   }  0.41  &               &  $0.57$    & 0.063  & 1.020 & 0.977   &  0.055    & 0.992   &0.969 \\
                          &              &                  &           &               &                & 0.097  & 1.070 & 0.983   & 0.089    &0.976   & 0.977 \\
                          &              &                  &           &               &                & 0.268  & 0.791 & 0.927   &  0.257    & 0.821  & 0.918 \\
                          &              &                  &           &               &                & 0.143  & 0.960 & 0.962   &0.133   &  0.930 & 0.955\\
                          &              &                  &           &               &              &  0.117  &1.241  & 1.022   &  0.115          &  1.236          & 1.021 \\
                          &              &                  &           &               &                & 0.154  & 1.311 & 1.024      & 0.153           & 1.292           & 1.023 \\
                          &              &                  &           &               &                & 0.334   & 1.035  &0.982    &0.336            & 1.031           & 0.981\\
Si$_2$Ge$_2$ &               &                  & $-$0.52 &           &  \llap{$-$}0.50    & 0.202  & 1.195 & 1.009   &  0.201 &1.186  & 1.008  \\
Si$_2$Ge$_2$/Si$_3$Ge$_1$                        & 3.8796 & 12.8128  &            &  3.880      &                   &             &            &           &            &            &    \\
Si$_3$Ge$_1$ &              &                & { }0.44  &               & 0.47	   & 0.026  & 1.259 & 0.937    & 0.024   & 1.249  & 0.935 \\
Si$_3$Ge$_1$ &              &                  & $-$0.49 &               & \llap{$-$}0.48     & 0.090  & 1.459 & 0.993  &  0.087 &1.452    & 0.992 \\
Si$_3$Ge$_1$/Si$_4$                         & 3.8435 & 12.7039  &             &    3.844           &                    &            &            &             &             &            &    \\
Si$_4$               &             &                 &  {   }  0.45  &               &  0.46           & $-$0.137 & 1.501 & 0.892  &$-$0.139 & 1.487 & 0.891\\  \hline
\end{tabular}
\label{tab5}
\end{table*}

The lattice constants and strains found for short-period superlattices should be evaluated in comparison with results of calculations for heterostructures with thicker
layers. In fact, the elastic energy \cite{Kittel:2005:Book} of a superlattice of pseudomorphically grown hexagonal materials  has to be minimized. For thicker layers of Si$_x$Ge$_{1-x}$ and Si$_y$Ge$_{1-y}$,  with thickness $D(x)$ and $D(y)$, one
finds as the minimum condition for the common in-plane lattice constant:
\begin{equation}\label{eq5}
a=\frac{D(x)Y(x)a(x)+D(y)Y(y)a(y)}{D(x)Y(x)+D(y)Y(y)} \,,
\end{equation}
using the biaxial moduli $Y=C_{11}+C_{12}-2(C_{13})^2/C_{33}$ and the in-plane lattice constants $a(x)$ and $a(y)$.This condition defines biaxial
strain of the two materials forming the heterostructure,
\begin{eqnarray}\label{eq6}
\epsilon_\|(x)=\frac{D(y)Y(y)}{D(x)Y(x)+D(y)Y(y)}\frac{a(y)-a(x)}{a(x)}, \nonumber \\
\epsilon_\|(y)=\frac{D(x)Y(x)}{D(x)Y(x)+D(y)Y(y)}\frac{a(x)-a(y)}{a(y)}.
\end{eqnarray}
In the case of equal thicknesses $D(x)=D(y)$, the mismatch in the elastic properties $Y(x)\neq Y(y)$ destroys the almost (apart from sign) symmetric distribution
$\epsilon_\|(x)\approx -\epsilon_\|(y)$. If one layer is much thicker than the other one, e.g. $D(y)\gg D(x)$, it holds that $a=a(y)$,
$\epsilon_\|(x)=\frac{a(y)-a(x)}{a(x)}$ and $\epsilon_\|(y)=0$.
In the case $D(x)=D(y)$, employing the $a$ and $Y$ parameters from Table~\ref{tab1}, expressions \eqref{eq5} and \eqref{eq6}
deliver similar common lattice constants and biaxial strains in Table~\ref{tab5} as in the case of short-period superlattices. The common lattice constant of the heterostructure \eqref{eq5}
 and the strains on both sides of the interface \eqref{eq6} tend, indeed, to approach the symmetric case $a=\frac{1}{2}[a(x)+a(y)]$ and $\epsilon_\|(x)\approx -\epsilon_\|(y)$. 
  
\subsection{\label{sec4c} Band edges and confinement in strained heterostructures}

The \emph{ab initio} calculations of the atomic geometry of the (Si$_n$Ge$_{4-n}$)$_{1}$/(Si$_{n+1}$Ge$_{3-n}$)$_{1}$(0001) ($n=0,1,2,3$) superlattices can be also
combined with MBJLDA calculations of the electronic structure. However, the resulting band structures in {\bf k}-space
cannot be immediately related to the band edges $E_v$, $E_c$, and $E^{\rm ind}_c$ versus a real space coordinate along the layer stacking.
An approximate approach to the band-edge profiles through the heterointerface in $z$-direction is possible by calculating the local  site-projected
density of states (PDOS). The PDOS is estimated by projection of the DOS onto each atom and by plotting the band edges around the local fundamental
gap versus the $z$-coordinate of the atomic layer \cite{Yamasaki.Kaneta.ea:2001:PRB,Seino.Bechstedt.ea:2010:PRB,Rauch.Marques.ea:2020:JCTC}.
Results for the four studied superlattices,  however, indicate that the projection technique is not applicable. The superlattice layers are too
thin. They mainly consist of interface regions, so that nearly constant band edge positions on both sides of an interface cannot be determined.

\begin{figure}[h]
\includegraphics[width=16cm]{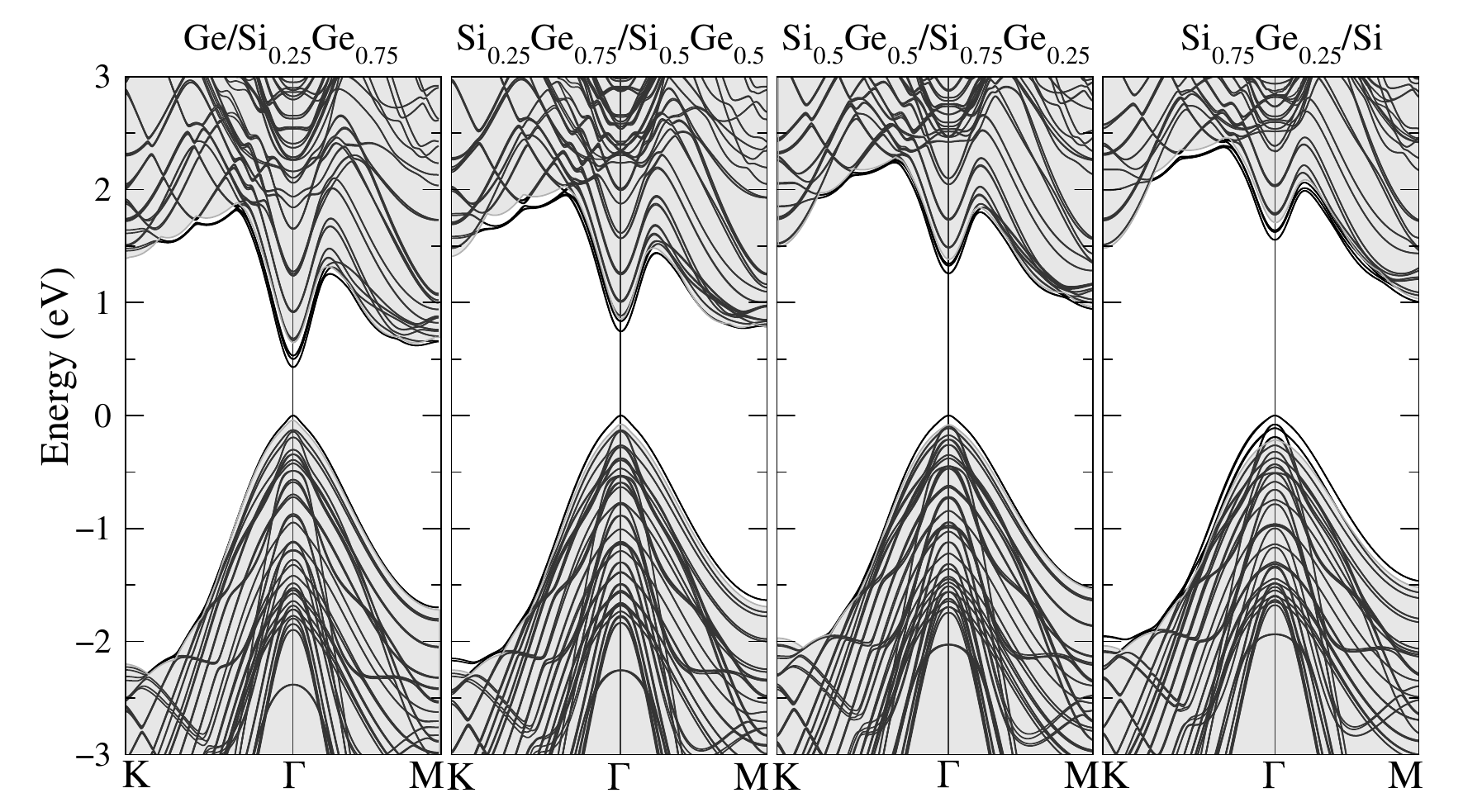}
\caption{Band structures of hexagonal (Si$_n$Ge$_{4-n}$)$_4$(Si$_{n+1}$Ge$_{3-n}$)$_4$(0001) superlattices. The VBM is taken as energy
zero. The shaded regions in the conduction and valence band regions indicate the allowed energy regions above and below the fundamental
gap of the more Si-rich alloy Si$_{n+1}$Ge$_{3-n}$ in its bulk form.}
\label{fig5}
\end{figure}

\begin{figure}[h]
\includegraphics[width=16cm]{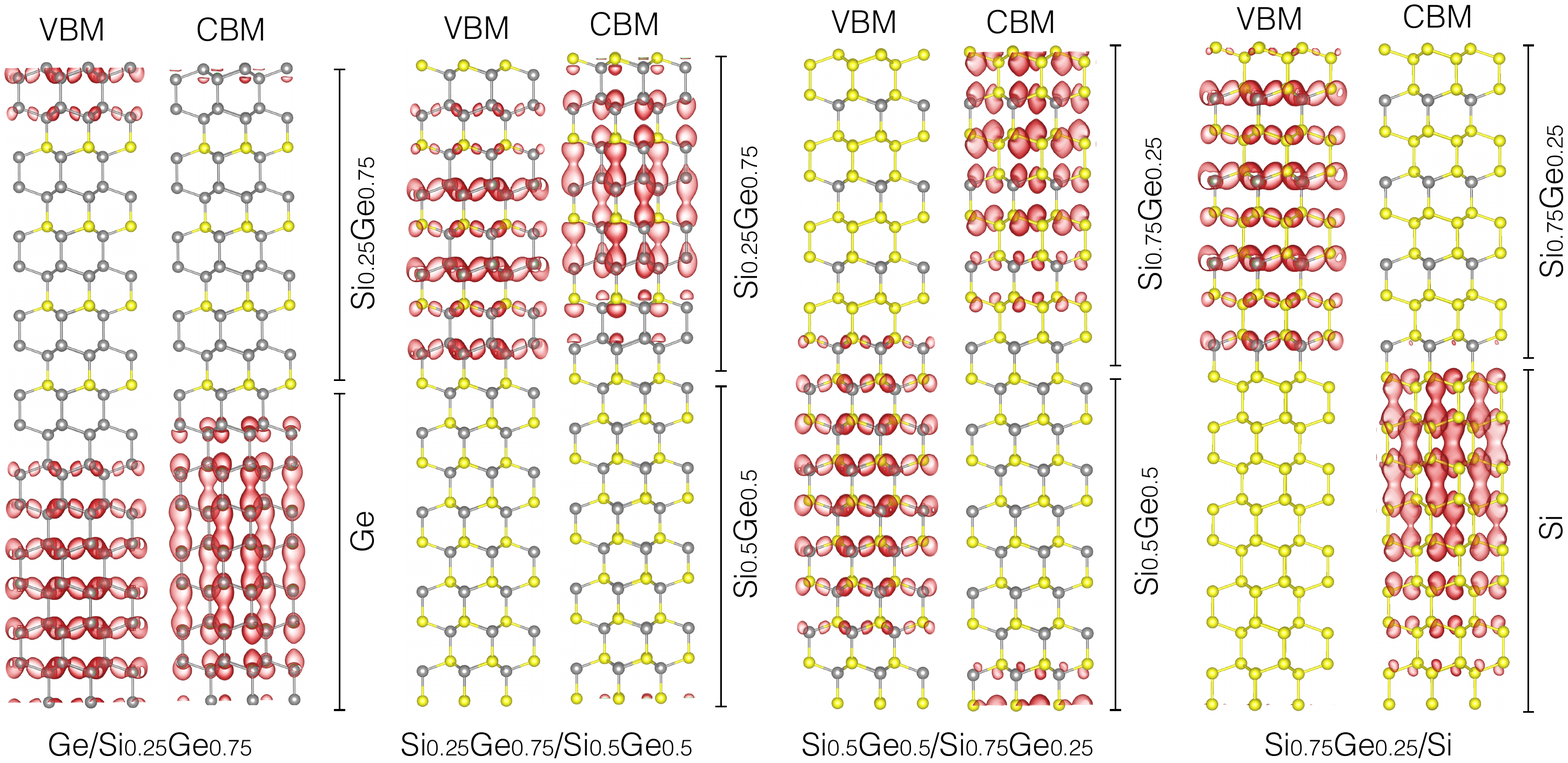} 
\caption{Wave-function squares of the lowest empty, i.e., CBM, and highest occupied, i.e., VBM, at $\Gamma$ together with two parallel
(11$\bar{2}$0) superlattice planes for hexagonal (Si$_n$Ge$_{4-n}$)$_4$(Si$_{n+1}$Ge$_{3-4}$)$_4$(0001) superlattices $(n=0,1,2,3)$.
Ge (Si) atoms are indicated by grey (yellow) circles. The 2H bond stacking is clearly visible.}
\label{fig6}
\end{figure}

For that reason we also studied superlattices with significantly thicker layers. A method described in detail in Ref.~\cite{Belabbes.Carvalho.ea:2011:PRB}
is applied. The electronic structure results are displayed in Figs.~\ref{fig5} and \ref{fig6} for superlattices with 32 atomic layers, more precisely
16 atomic layers on each side of an interface, in one superlattice unit cell. In other words hexagonal 
(Si$_{n}$Ge$_{4-n}$)$_4$(Si$_{n+1}$Ge$_{3-n}$)$_4$(0001) superlattices with $n=0,1,2,3$ are studied. The resulting band structures are shown
in Fig.~\ref{fig5} together with a background that illustrates the band regions of the projected band structure of the hexagonal barrier materials
consisting of Si$_{n+1}$Ge$_{3-n}$ four-atom unit cells. 
The BP alignment is applied to align the superlattice and the projected band structures in each panel of Fig.~\ref{fig5}. The VBM of the superlattices is
used as energy zero for all superlattices. 

The direct gaps at $\Gamma$ of the four superlattices are 0.43 ($n=0$), 0.74 ($n=1$), 1.26 ($n=2$), and 1.55 ($n=3$) eV larger than the direct gaps 
of the hexagonal Si$_n$Ge$_{4-n}$ materials given in Table~\ref{tab3}. This observation indicates the presence of quantum confinement of electrons and holes at $\Gamma$
in the Ge-richer superlattice layers, or type-I heterostructure behavior of carriers in these regions, in agreement with the natural band lineups of Fig.~\ref{fig3}. 
For the indirect gaps, we report a gap increase of 0.62 ($n=0$), 0.77 $(n=1)$, 0.94 ($n=2$), and 1.00 ($n=3$) eV for the CBM at the M point of the superlattice BZ, showing that confinement is less pronounced. Electron
 confinement is present in the direct-gap $n=0$ and $n=1$ superlattices, while the indirect-gap $n=2$ and $n=3$ superlattices tend to smaller values, clearly visible 
 for the (Si$_3$Ge$_1$)$_4$(Si$_4$)$_4$(0001) superlattice, where the projected bulk Si band structure is below the superlattice CBM at M. These findings are in agreement
  with Fig.~\ref{fig3}, which clearly suggests hole confinement in the Ge-rich regions when the MBJLDA  approach is applied. 
We observe that the energy distance between the band extrema of the superlattice bands and the projected band structure are somewhat smaller than the values 
found for the natural band discontinuities in Fig.~\ref{fig5} and Table~\ref{tab3}. The hole confinement in the Si-poorer alloy regions of all superlattices is visible. 
From the distance to the shaded region one may conclude that the deepness of the corresponding hole quantum well increases with rising Si composition in 
the superlattice. For electrons the situation is more complex. For Si-poor compositions the superlattices are direct semiconductors and localized electron states appear at $\Gamma$. 
This behavior at $\Gamma$ continues in Si-richer superlattices. However, the true conduction band minima appear at M in the superlattice BZ. The confinement
of electrons at $\Gamma$ in the Ge-richer layers is hardly visible. Summarizing, the band structures of the Ge-rich superlattices clearly show a type-I heterobehavior
using MBJLDA bands, as suggested by the natural band lineups in Fig.~\ref{fig3}. The situation for lower Ge contents is less clear. 

This tendency for type-I heterobehavior is not only indicated by the lowest conduction band and highest valence band in the fundamental gap of the barrier material
in the Ge-richer direct-gap superlattices in Fig.~\ref{fig5}. A pronounced type-I (type-II) heterostructure behavior of the Ge-rich (Si-rich) superlattices is also
demonstrated by the wave function squares of the superlattice Bloch functions of the lowest conduction and highest valence band, as shown in Fig.~\ref{fig6}.
In Ge-richer superlattices both wave functions, for electrons and holes at $\Gamma$, are localized in the Ge-rich layer of the
superlattice structure, clearly representing a type-I behavior. One may speak about a multi-quantum well structure with  electron
and hole wells in the Ge-rich layers, while Si-richer layers form barriers for both carrier types. There is a complete change in Fig.~\ref{fig6}
for the Si-richer superlattices. Hole and electron wave functions are now localized in different regions of the superlattice.
A type-II heterojunction behavior is suggested by the wave function localization. The holes remain localized in the Si-poorer layers, while electrons are more likely to be in Si-richer layers.

\begin{figure}[h]
\includegraphics[width=16cm]{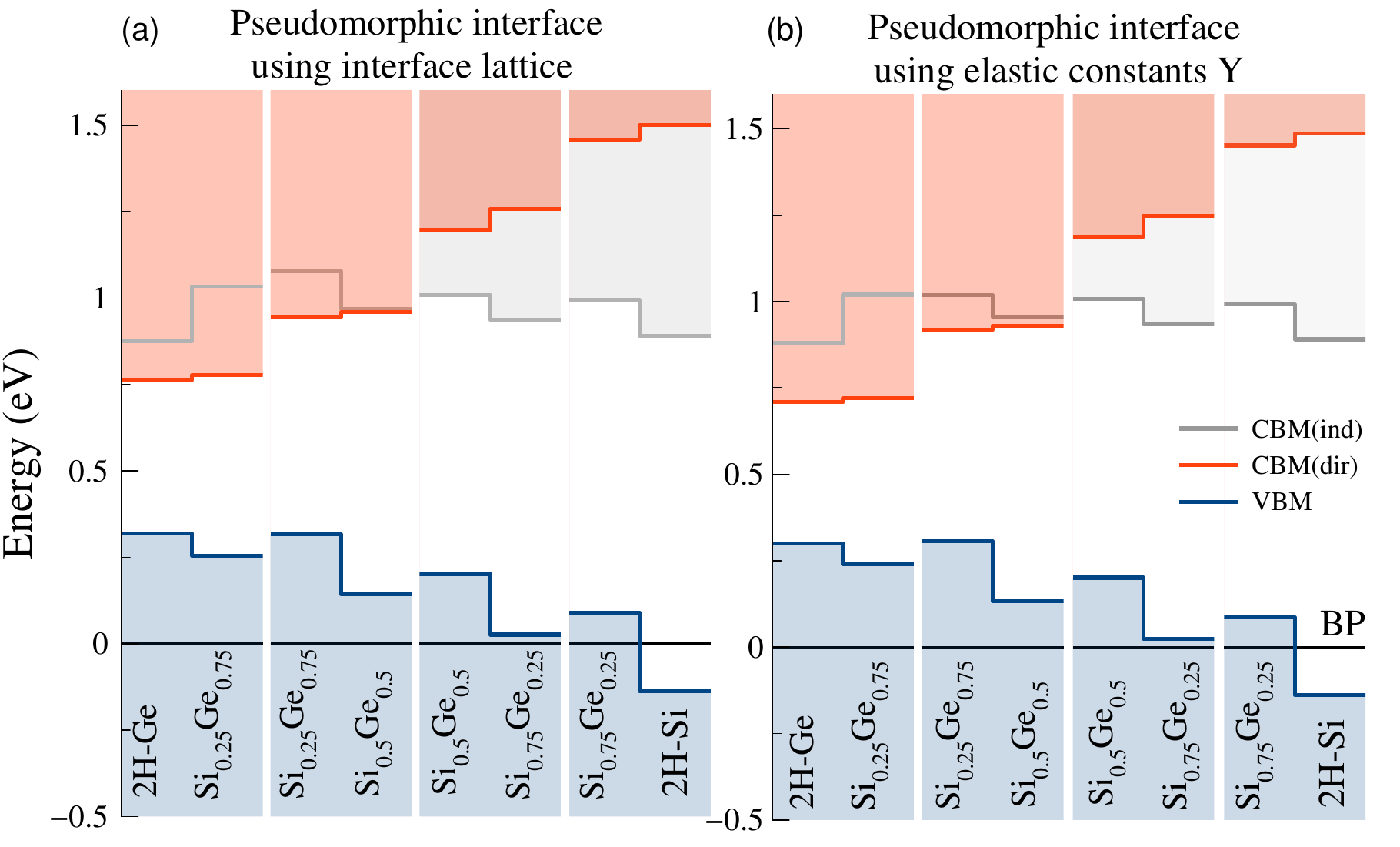} 
\caption{Band lineups at the interfaces of the hexagonal Si$_n$Ge$_{4-n}$/Si$_{n+1}$Ge$_{3-n}$ ($n=0,1,2,3$) heterostructures as calculated by means of
expression  \eqref{eq4} and the energies and deformation potentials in Table~\ref{tab4} using the biaxial strains \eqref{eq3} from (a) the superlattice and (b) the macroscopic
approach in \eqref{eq6}. The resulting strain values are listed in Table~\ref{tab5}.}
\label{fig7}
\end{figure}

\begin{figure}[h]
\includegraphics[width=13cm]{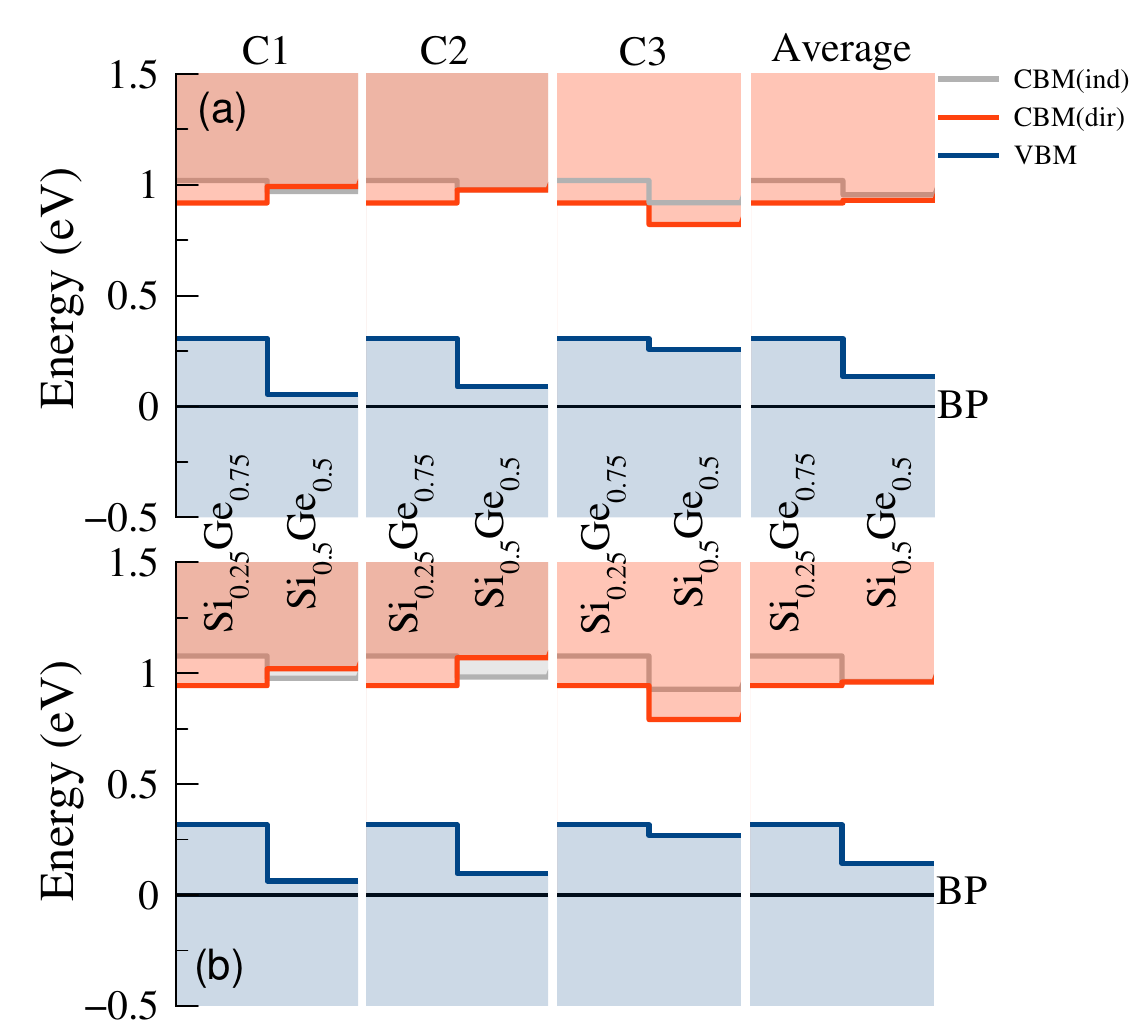} 
\caption{Zoom to heterostructures with layer compositions $x=0.25$ or $y=0.5$. Besides the average result already displayed in Fig.~\ref{fig7} also
band lineups for the defined atomic configurations C1, C2 and C3 of the Si$_2$Ge$_2$ cells are plotted. The different biaxial strains used in (a) and (b) are estimated
as described in Fig.~\ref{fig7}.}
\label{fig8}
\end{figure}

In order to determine the pure effect of the biaxial strain on the band lineups between the two different materials,
 biaxial strains are extracted for short-period superlattices with layers of equal thickness in the unit cell and listed in Table~\ref{tab5}. 
These strains \eqref{eq3}, together with the deformation potentials of Table~\ref{tab4}, lead to band edge positions \eqref{eq4} with
respect to the BP as energy zero. They lead to the plots of Fig.~\ref{fig7} for the band positions and band discontinuities. We note that in Fig.~\ref{fig7}(a) the strains directly
extracted from the \emph{ab initio} optimized geometries of the short-period superlattice are used, while the biaxial strains \eqref{eq6} estimated by means of the elastic moduli $Y$ and the in-plane
lattice constants $a$ of the unstrained alloys, both in Table~\ref{tab1}, are applied in Fig.~\ref{fig7}(b). Qualitatively, both Figs.~\ref{fig7}(a) and
\ref{fig7}(b) show the same band lineups with pronounced quantum wells for holes in the Ge-rich material layer. However, in the light of the goal to fabricate hexagonal Si$_x$Ge$_{1-x}$ heterostructures, which are suitable for active optoelectronic devices, Fig.~\ref{fig7} exhibits a somewhat more pronounced tendency for hole confinement compared to the natural band lineups in Fig.~\ref{fig3}. The conduction band offsets guaranteeing a type-I heterostructure are generally reduced and all the heterojunctions with more silicon mixed-in into the well,
as well as barrier material, exhibit a clear tendency to become indirect semiconductors, with the lowest conduction band minimum located at
the LM line in {\bf k}-space or M point in the superlattice BZ. Only the hexagonal Ge/Si$_{0.25}$Ge$_{0.75}$ heterojunction, and perhaps also the Si$_{0.25}$Ge$_{0.75}$/Si$_{0.5}$Ge$_{0.5}$
interface, represents a type-I heterostructure, however, with two deficiencies. (i) The barrier for the electrons and, therefore, their quantization in the pure 2H-Ge or Si$_{0.25}$Ge$_{0.75}$ layer is much smaller compared to the result for natural alignment in Fig.~\ref{fig3}. (ii) The lowest-energy optical transitions inside 2H-Ge still possesses only a small oscillator strength
\cite{Belabbes.Bechstedt.ea:2021:PSSRRL}. In order to improve the situation for the construction of a heterostructure laser based on a
hexagonal Si$_x$Ge$_{1-x}$/Si$_y$Ge$_{1-y}$ alloy system, one has to find a way to reduce the compressive biaxial strain in the Ge-richer
layer. One option to do so is to increase the thickness $D(x)$ compared to $D(y)$. Then, according to \eqref{eq6} the
compressive strain $|\epsilon_\|(x)|$ can be reduced in comparison to the tensile strain $|\epsilon_\|(y)|$. Together with the 
condition to have a reasonable strength of the optical transitions \cite{Belabbes.Bechstedt.ea:2021:PSSRRL}, the heterostructure 
Si$_{0.25}$Ge$_{0.75}$/Si$_{0.5}$Ge$_{0.5}$ could be a promising system if, indeed, a larger thickness of the Ge-rich layer is reached. In Fig.~\ref{fig7} average values are presented in the Si$_2$Ge$_2$ case. Figure~\ref{fig8} shows the band lineups for a situation where in the heterostructures the alloy with $y=0.5$ is realized by only one 
configuration, C1, C2, or C3 (see Fig.~\ref{fig1}). The hole wells appear to be in the Ge-rich regions independent of the atomic configuration. However, the position of the CBM
varies with the atomic distribution corresponding to the Si-richer layer. A type-I (with C1 and C2) or type-II (with C3) heterostructure character appears in dependence of the actual atomic
configuration.

\section{\label{sec5}Summary and conclusions}

The electronic properties of heterostructures made of hexagonal Si$_x$Ge$_{1-x}$ alloys have been studied by means of \emph{ab initio}
calculations of atomic geometries, based on density functional theory and using approximate quasiparticle approaches for the band structures. Two different alignment procedures to
construct natural band lineups are tested, together with the direct calculation of various hexagonal unit cells to describe alloys and heterojunctions. We applied elastic theory to model the
influence of biaxial strain, in particular in the case of pseudomorphically grown heterosystems.

The natural band lineups have been investigated comparing the branch point and vacuum level alignments. Moreover, the influence of the underlying approximate quasiparticle method, namely the MBJLDA or HSE06 functionals, is considered. We can conclude that the selection of one of these approximation has a negligible effect in terms of the general accuracy
of the band offsets $\Delta E_c$ and $\Delta E_v$. The different alignment methods lead also to similar results, so that we decided to choose the branch point alignment, that only requires bulk calculations. Calculations for diamond Si and Ge allowed to validate the values of the branch point
energies and the natural band discontinuities, comparing with experiemnts and previous calculations. Consequences of the negative branch points in Ge-rich
alloys have been predicted for carrier accumulation at hexagonal Ge and alloy surfaces.

The true band offsets appearing at almost pseudomorphic heterointerfaces have been investigated for different thicknesses of the alloy
layers composing the heterojunction. The limit of thin layers has been studied within a superlattice approach. It gives rise to compressive
(tensile) biaxial strains of order of $\pm0.5$~\% in the Ge-richer (less Ge-rich) alloys. However, for larger thicknesses,
but always in the range of validity of elastic theory, similar strains result from the minimization of the elastic energy of the junction. The strain profiles
with compressive strain at the Ge-richer side and tensile strain at the Si-richer side tend to deepen the hole
quantum wells, and flatten the electron quantum wells in the Ge-poorer layers. Also the type-I character of the heterostructures
is reduced. This tendency is accompanied by a transition of the heterostructure from a direct semiconductor to an indirect one with rising Si composition. These results are confirmed by electronic structure calculations of the band structures and the wave
functions, performed for superlattices with thicker layers. Consequently, we propose to achieve a reduction of the compressive strain 
at the Ge-rich side to make the pseudomorphic heterojunctions suitable for active optoelectronic applications.

\section*{Acknowledgements}

We acknowledge financial support from the H2020-FETOpen project OptoSilicon (grant agreement No. 964191).

\bibliography{Bandlineup.bib}

 \newpage
 \section*{\label{figcap}Figure Captions}

\end{document}